\documentclass[aps,prl,superscriptaddress,reprint,showpacs,floatfix]{revtex4-1}
\usepackage{graphicx}
\usepackage{dcolumn}
\usepackage{bm}
\usepackage{xcolor}
\usepackage{relsize}
\usepackage{amsmath}
\usepackage{braket}
\usepackage{gensymb}
\usepackage{url}
\usepackage{footmisc}
\usepackage[utf8]{inputenc}
\begin{document}

\preprint{APS/123-QED}

\title{Exciton Polaritons in a Two-Dimensional Lieb Lattice with Spin-Orbit Coupling}

\author{C. E. Whittaker}
\email{cewhittaker1@sheffield.ac.uk}
\affiliation{Department of Physics and Astronomy, University of Sheffield, Sheffield S3 7RH, United Kingdom}%
\author{E. Cancellieri}
\affiliation{Department of Physics and Astronomy, University of Sheffield, Sheffield S3 7RH, United Kingdom}%
\author{P. M. Walker}
\affiliation{Department of Physics and Astronomy, University of Sheffield, Sheffield S3 7RH, United Kingdom}%
\author{D. R. Gulevich}
\affiliation{ITMO University, St. Petersburg 197101, Russia}%
\author{H. Schomerus}
\affiliation{Department of Physics, Lancaster University, Lancaster LA1 4YB, UK}%
\author{D. Vaitiekus}
\affiliation{Department of Physics and Astronomy, University of Sheffield, Sheffield S3 7RH, United Kingdom}%
\author{B. Royall}
\affiliation{Department of Physics and Astronomy, University of Sheffield, Sheffield S3 7RH, United Kingdom}%
\author{D. M. Whittaker}
\affiliation{Department of Physics and Astronomy, University of Sheffield, Sheffield S3 7RH, United Kingdom}%
\author{E. Clarke}
\affiliation{EPSRC National Centre for III-V Technologies, University of Sheffield, Sheffield S1 3JD, United Kingdom}
\author{I. V. Iorsh}
\affiliation{ITMO University, St. Petersburg 197101, Russia}
\author{I. A. Shelykh}
\affiliation{ITMO University, St. Petersburg 197101, Russia}
\affiliation{Science Institute, University of Iceland, Dunhagi 3, IS-107, Reykjavik, Iceland}
\author{M. S. Skolnick}
\affiliation{Department of Physics and Astronomy, University of Sheffield, Sheffield S3 7RH, United Kingdom}
\affiliation{ITMO University, St. Petersburg 197101, Russia}%
\author{D. N. Krizhanovskii}
\email{d.krizhanovskii@sheffield.ac.uk}
\affiliation{Department of Physics and Astronomy, University of Sheffield, Sheffield S3 7RH, United Kingdom}
\affiliation{ITMO University, St. Petersburg 197101, Russia}%

\date{\today}

\begin{abstract}
We study exciton-polaritons in a two-dimensional Lieb lattice of micropillars. The energy spectrum of the system features two flat bands formed from $S$ and $P_{x,y}$ photonic orbitals, into which we trigger bosonic condensation under high power excitation. The symmetry of the orbital wave functions combined with photonic spin-orbit coupling gives rise to emission patterns with pseudospin texture in the flat band condensates. Our work shows the potential of polariton lattices for emulating flat band Hamiltonians with spin-orbit coupling, orbital degrees of freedom and interactions.

\end{abstract}

\maketitle

Two-dimensional lattices with flat energy bands attract keen research interest as a platform to study exotic many-body effects including itinerant ferromagnetism \cite{PhysRevA.82.053618}, Wigner crystallization \cite{PhysRevLett.99.070401} and fractional quantum Hall phases \cite{PhysRevLett.107.146803}. A notable example of a flat-band system is the Lieb lattice \cite{PhysRevLett.62.1201}, a decorated square lattice found in nature in the cuprates exhibiting high-$T_{c}$ superconductivity \cite{Keimer} and studied extensively in recent years for its topologically nontrivial phases \cite{PhysRevB.82.085310,PhysRevA.83.063601,PhysRevB.86.195129,doi:10.1142/S021797921330017X,1367-2630-17-5-055016,2015arXiv151007239K,PhysRevB.92.235106,PhysRevA.93.043611,1674-1056-25-6-067204,WangR2016,PhysRevLett.117.163001}. In bosonic systems, models of particles in two-dimensional Lieb lattice potentials with flat energy bands are a highly valuable tool for researchers, having recently been experimentally realized in photonic waveguide arrays \cite{1367-2630-16-6-063061,PhysRevLett.114.245504,PhysRevLett.116.183902,2053-1583-4-2-025008} and ultracold atoms in optical lattices \cite{Taiee1500854}. Particularly fascinating prospects which remain unexplored in Lieb lattice models are many-body interactions, spin-orbit coupling (SOC) terms and orbital structure. With such features the flat bands are predicted to support nonlinear compactons \cite{PhysRevA.93.043847,PhysRevB.94.144302} and interaction-induced topological phases \cite{PhysRevLett.117.163001}. More generally, these lattices allow one to study the interplay between fundamental nonlinear, spin and orbital phenomena in a topological system. 

Exciton-polariton (polariton) gases confined in lattice potentials have recently emerged as an attractive candidate for emulating nonlinear lattice Hamiltonians \cite{Amo2016934}. Microcavity polaritons are the mixed light-matter eigenmodes characterized by a small effective mass, allowing both quasi-equilibrium and nonequilibrium Bose-Einstein condensation at elevated temperatures \cite{KasprzakJ2006,Balili1007,PhysRevLett.101.067404,PhysRevB.80.045317}. Giant exciton-mediated Kerr nonlinearity, which is 3--4 orders of magnitude larger than light-matter systems in the weak-coupling regime \cite{Walker2015}, has enabled the observation of ultra-low power solitons \cite{SICH2016908} and vortices \cite{proukakis_snoke_littlewood_2017}, and more recently driven-dissipative phase transitions associated with quantum fluctuations \cite{PhysRevLett.118.247402,PhysRevX.7.031033,1707.01837}. In polariton systems, straightforward optical techniques can be used to create 
interacting scalar and spinor boson gases in highly tunable lattice geometries, which can be engineered through modulation of the photonic \cite{KimNY2011,PhysRevLett.112.116402,1367-2630-17-2-023001} or excitonic \cite{PhysRevLett.105.116402,PhysRevB.87.155423,PhysRevB.86.100301,PhysRevB.87.155423,TosiG2012} potential landscape. Furthermore, the spatial, spectral and pseudospin (polarization) properties of the polaritonic wave functions are directly accessible due to the finite cavity photon lifetime. The inherently nonequilibrium nature of polariton gases also means that higher energy orbital bands, formed from spatially-anisotropic modes, are readily populated, as was recently demonstrated in a honeycomb lattice \cite{PhysRevLett.118.107403}.

One intriguing property of polaritons in lattices is polarization-dependent tunneling \cite{PhysRevX.5.011034}, inherited mostly from the photonic component and enhanced by TE-TM splitting. It is formally analogous to SOC \cite{PhysRevX.5.011034,PhysRevLett.115.246401,Solnyshkov2016920} inducing a $k$-dependent effective magnetic field acting on polariton pseudospin. The rich variety of polarization phenomena exhibited by polaritons in both noninteracting and interacting regimes \cite{0268-1242-25-1-013001} remains unexplored in two-dimensional periodic potentials, and is inaccessible in the inherently asymmetric one-dimensional case \cite{PhysRevLett.116.066402}. 

In this Letter we study a two-dimensional (2D) array of coupled micropillars arranged in a Lieb lattice. The crystal structure comprises three square sublattices (denoted $A$, $B$ and $C$) each contributing one atom to the unit cell [Fig. \ref{fig1}(a)]. This lattice topology, in which the sites on different sublattices have different connectivity, results in localized states residing on dispersionless energy bands \cite{PhysRevB.34.5208}. Here we explore the bands formed by evanescent coupling of both the ground and first excited states of the pillars, which are 2D photonic orbitals with $S$ and $P$ like wave functions. We excite the system quasi-resonantly to optically load polaritons into the periodic potential, triggering condensation into three separate modes of the lattice – $S$ and $P$ flat (non-bonding) bands and the maxima of the $S$ anti-bonding (AB) band. Resolving the near-field emission in energy and polarization above the threshold for polariton condensation we see that the flat band condensates show novel pseudospin textures arising from a polarization-dependent hopping energy, which acts as SOC for polaritons. Significant variation in the emission energy across real space (fragmentation) can be seen in the flat-band condensates, which we show arises due to the effect of many-body interactions, since the kinetic energy scale is quenched. This contrasts with the condensates formed on the $S$ AB (dispersive) band which emit at strictly one energy. 

\begin{figure}
\center
\includegraphics[scale=1]{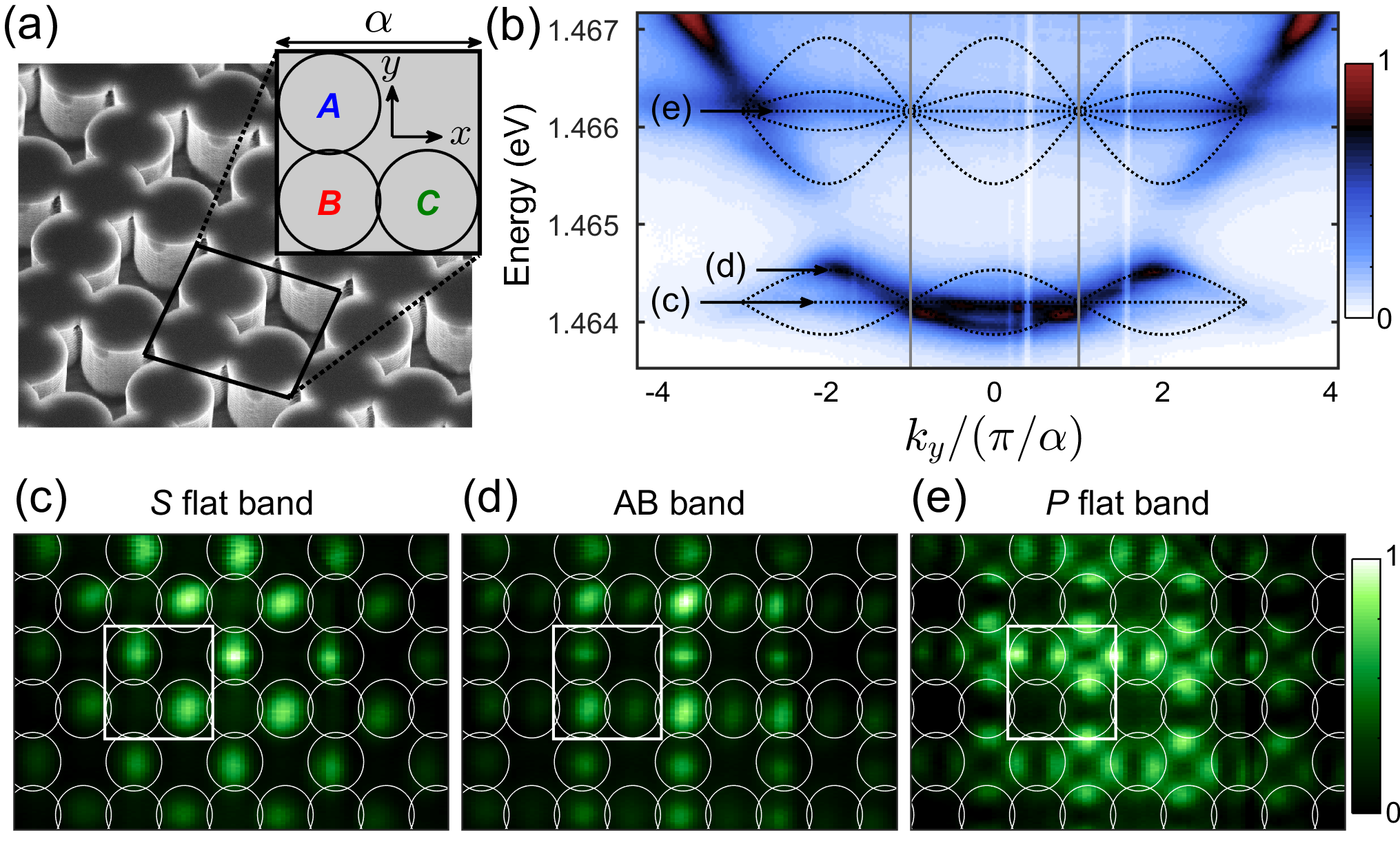}
\center
\vspace{-0.4cm}
\caption{(a) Scanning electron microscope image of a section of the 2D Lieb lattice. The enlarged image shows a schematic diagram of 1 unit cell and the 3 sublattices. (b) Lattice emission in energy-momentum space measured under low-power nonresonant excitation. White dotted lines correspond to bands calculated from the tight-binding model. (c)-(e) Emission in real space at the energies of the $S$ flat band (c), the AB band maxima (d) and the $P$ flat band (e). The white squares correspond to one unit cell as shown in (a) and the corresponding energies are shown by white arrows in (b).}
\label{fig1}
\end{figure}

Our 2D Lieb lattice consists of AlGaAs/GaAs micropillars of 3 $\mu$m diameter and a separation of 2.9 $\mu$m. The lattice periodicity is $\alpha = 5.8$ $\mu$m. Further details about the sample and experimental methods can be found in Ref. \footnote{See supplementary material at [url] for further details about the sample, experiment and theoretical models, which includes Ref. \cite{PhysRevB.76.201305}}. The single-particle band structure of our Lieb lattice at $k_{x} = \pi/\alpha$ is displayed in Fig. 1(b). It shows the energy bands associated with the two lowest energy pillar modes, $S$ and $P$ (comprising degenerate $P_{x}$ and $P_{y}$) orbitals which have bare energies of around 1.4642 eV and 1.4662 eV respectively. Photonic coupling results in $S$ and $P$-type flat bands [(c) and (e) in Fig. 1(b)] separated by 2 meV and a forbidden energy gap of approximately 0.8 meV between $S$ and $P$ dispersive bands. The white dotted lines in Fig.\ref{fig1}(b) correspond to the dispersion curves calculated from our tight-binding (TB) model \cite{Note1}. Experimentally the emission from some of the expected folded branches of the dispersion appears almost absent. For example, emission from the $S$ AB band [(d) in Fig.\ref{fig1}(b)] is suppressed within the first Brillouin zone (delimited by vertical lines) whilst the $S$ bonding band is suppressed outside. This effect can be attributed to a combination of far-field destructive interference and varying lifetimes (due to relaxation and losses) of different modes. This effect is well known in honeycomb lattices \cite{PhysRevLett.112.116402,PhysRevB.51.13614}, and in Ref. \cite{Note1} we solve the 2D Schr\"{o}dinger equation for a periodic potential to
confirm that this is also the case for the Lieb lattice. 

In Fig. \ref{fig1}(c) and (e) we show the real space distribution of polariton emission intensity of the $S$ and $P$ flat bands, constructed by scanning the emission across the spectrometer slit and piecing together the energy-resolved slices. 
In both cases there is highly suppressed emission from the $B$ sublattice, characteristic of flat bands, indicating that polaritons are highly localized on $A$ and $C$ sublattices. This results from destructive wave interference of the $A$ and $C$ sublattice linear eigenmodes due to the local lattice symmetries \cite{PhysRevB.87.125428}. Interestingly, for the $P$ band, we see that emission from $P_{x}$ orbitals dominates on the $A$ sites and $P_{y}$ orbitals for the $C$ sites (the subscript denotes the axis to which the two lobes lie parallel). Since orthogonal $P$ orbitals do not interfere with each other, the absence of emission from $B$ sites must arise from destructive interference of like $P$ orbitals. A qualitative explanation is that the difference in the tunneling energies for spatially anisotropic modes with orthogonal orientations ($\sigma$ and $\pi$ bonding) is offset by a difference in the orbital populations on the two sites maintaining the destructive interference necessary for flat band formation. We expand this argument including polarization later in the text. In contrast to the two flat bands, at the energy maxima of the dispersive AB mode [Fig. \ref{fig1}(d)] the polaritons are delocalized across all 3 sublattices as is usually expected for the linear eigenmodes of a periodic potential. 

\begin{figure}
\center
\includegraphics[scale=1]{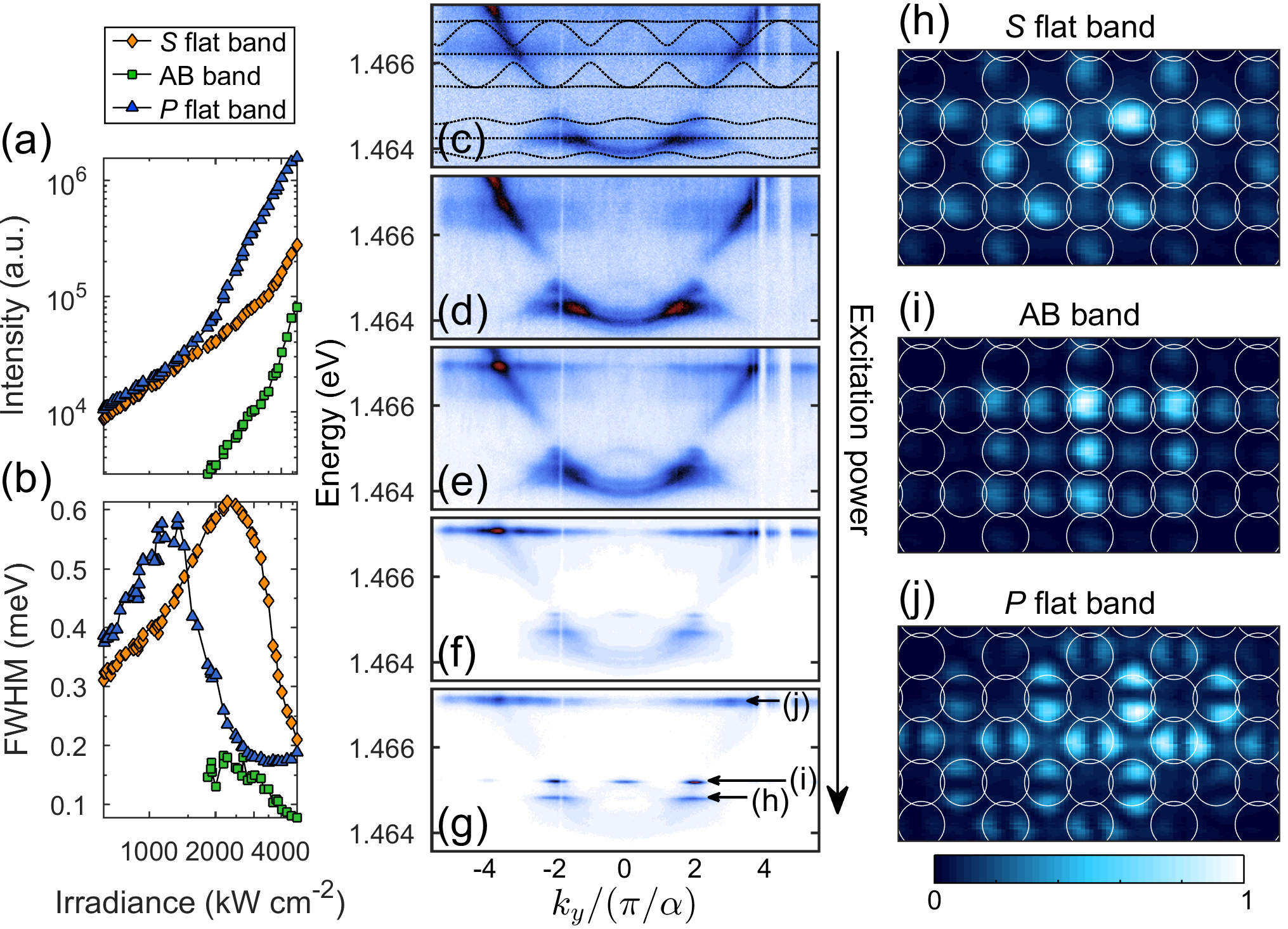}
\center
\vspace{-0.4cm}
\caption{(a) Peak intensity and (b) full-width half-maximum of the three lattice modes as a function of the sample irradiance in the vicinity of the condensation thresholds. (c)-(g) The far-field emission with increasing irradiance, with condensation occurring in (f) and (g). The sample irradiances are 620 (c), 1360 (d), 2180 (e), 4030 (f) and 5630 kW cm$^{-2}$ (g). The color scale is the same as that of Fig. \ref{fig1}(b). (h)-(j) Real space images of the lattice condensates.}
\label{intensity_dependence}
\end{figure}

In order to study our system in the nonlinear kinetic condensation regime we tune the pump laser to 843 nm, resonant with high energy states of the lower polaritonic bands (detuned roughly -1 meV from the exciton) where a broad continuum of high energy pillar modes exists \cite{Note1}. Through this channel we resonantly inject polaritons into the lattice at normal incidence, using high irradiances to create large populations of interacting polaritons. We use a large pump spot ($\sim$25 $\mu$m) which excites around 15 unit cells of the lattice. In Fig. \ref{intensity_dependence} we show the evolution of our system with sample irradiance. Figs. \ref{intensity_dependence}(c)-(g) show the momentum space emission at $k_{x}$ = 0. Beyond a critical pumping intensity, macroscopic populations of particles begin to accumulate in the $P$ flat band as evidenced by a superlinear increase in the emission intensity [Fig. \ref{intensity_dependence}(a)] and narrowing of the linewidth [Fig. \ref{intensity_dependence}(b)] which signifies increased temporal coherence \cite{PhysRevLett.97.097402,PhysRevLett.101.067404}. A similar condensation process is seen at slightly higher pumping intensities for the $S$ AB band maxima, which move into the spectral gap, and the $S$ flat band. As can be seen in Figs. \ref{intensity_dependence}(f),(g) these co-existing condensates dominate the normalized PL spectra above threshold. In the $S$ AB band, condensates are formed in the negative effective mass states with an energy residing in the forbidden gap, which is reminiscent of the gap solitons previously reported as nonlinear solutions in similarly shallow periodic potentials \cite{TaneseD2013,PhysRevLett.111.146401}. The real space distributions of the three condensate modes are shown in Figs. \ref{intensity_dependence}(h)-(j). The dark $B$ sites observed for the flat band cases confirm that the condensates indeed reside on highly nondispersive energy bands, in contrast to the condensates formed at the maxima of the dispersive AB band. 

\begin{figure}[b]
\center
\includegraphics[scale=1]{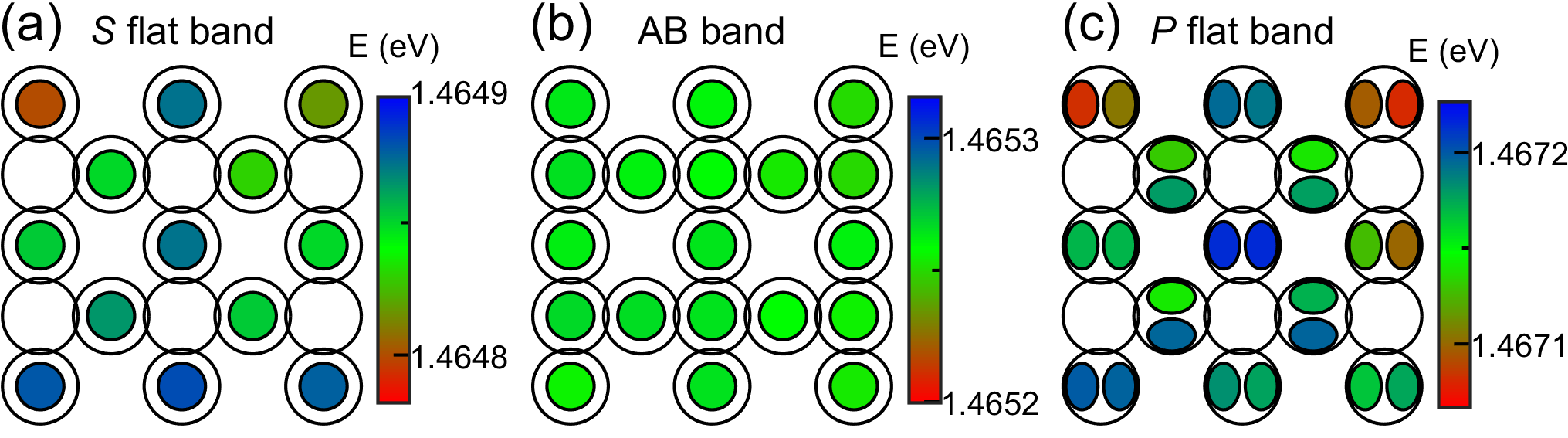}
\center
\vspace{-0.5cm}
\caption{(a) Color maps showing spatial energy variation of the $S$ flat band (a), AB band (b) and $P$ flat band (c) condensate emission constructed from experimental data above threshold.}
\label{fig_fragmentation}
\end{figure}

Above the condensation threshold, the high density of polaritons leads to a sizable mean-field interaction energy due to Coulomb interactions between polaritons residing in the condensates as well as interactions of condensed polaritons with the highly populated resonantly pumped states. For the case of flat bands, the kinetic energy scale is quenched due to an infinite effective mass, so energy renormalization is non-trivial and cannot be treated as a perturbation \cite{PhysRevB.82.184502}. With no kinetic energy to counterbalance the local nonlinear interaction energy a fragmentation of the condensates into localized modes emitting at slightly different energies is observed. 
Conversely, the AB band polaritons acquire kinetic energy when they propagate from high to low density regions compensating the low potential energy in low density regions, resulting in a homogeneous emission energy across the lattice in real space. Spatial maps constructed from experimental data above threshold are shown in Fig. \ref{fig_fragmentation} and demonstrate the degree of spectral variation of the three condensed modes, which is vanishing for the AB band but pronounced for the flat bands. In Ref. \cite{Note1} we analyze the relation between the population and energy across the condensates and show correlations which provide further evidence for the strong influence of many-body interactions in the fragmentation of the flat energy bands.

\begin{figure}
\center
\includegraphics[scale=1]{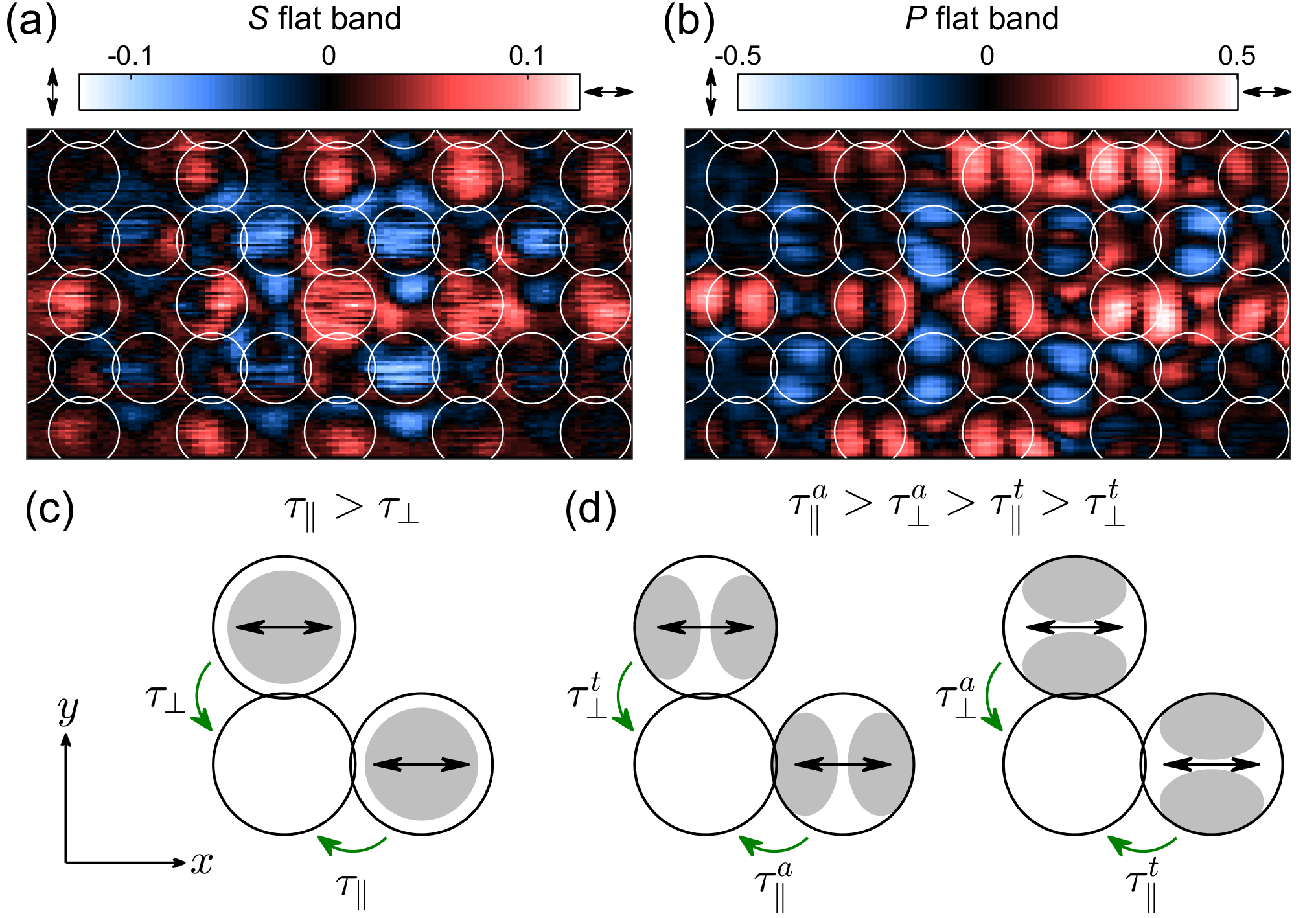}
\center
\vspace{-0.5cm}
\caption{$S_{1}$ linear Stokes parameter of the real space emission at the energies of the $S$ (a) and $P$ (b) flat band. The color scale is linear with red (blue) representing H (V) polarization, depicted by arrows either side of the color bar. (c),(d) Schematic of the nearest neighbor hopping processes for $S$, $P_{x}$ and $P_{y}$ orbitals, in the case of polarization-dependent hopping probabilities.}
\label{fig_polarisation}
\end{figure}

So far we have studied the spatial and spectral properties of polaritons in both dispersive and flat bands, demonstrating bosonic condensation and analyzing the effect of non-perturbative many-body interactions. Now we consider the pseudospin degree of freedom by resolving the lattice emission in polarization. In Figs. \ref{fig_polarisation}(a) and (b) we plot the Stokes linear polarization parameter $S_{1}$ for the flat band condensates, with $S_{1} = (I_{H}-I_{V})/(I_{H}+I_{V})$  where $I_{H}$ and $I_{V}$ are the intensities of the emitted light measured in the horizontal (0\degree) and vertical (90\degree) bases respectively. Ordered pseudospin textures extended across several unit cells can be seen. The finite spatial extent of the pump spot and the intensity-dependent blueshift associated with its Gaussian profile limits the size of the observed patterns. 

In order to explain these polarization patterns one needs to consider the two following features of Lieb lattices and Bragg-cavity polariton systems. First, in Lieb lattices the eigenmodes associated with the flat bands are non-spreading modes characterized by having zero population on the $B$ sites. This characteristic feature is due to the destructive interference of particles tunneling to $B$ sites from the neighboring $A$ and $C$ sites \cite{PhysRevB.87.125428}. Clearly, in the case of different tunneling probabilities, the destructive interference can occur only if the neighboring sites have different populations, since the number of particles tunneling from one site to another is proportional both to the tunneling probability and to the number of particles on an initial site.

Second, in lattices formed from Bragg cavities, the particles’ tunneling probability from one pillar to another has been experimentally observed \cite{PhysRevX.5.011034} and theoretically discussed \cite{PhysRevLett.114.026803} to be polarization-dependent, leading to effective spin-orbit coupling. In particular, it has been shown that the tunneling probability ($\tau_{\parallel}$) of particles having polarization parallel to the propagation direction is higher than the tunneling probability ($\tau_{\perp}$) of particles having polarization perpendicular to the propagation direction:
\begin{equation}
\label{tunnelling_relation}
\tau_{\parallel}>\tau_{\perp}.
\end{equation}

In the case of the Lieb lattice this means that horizontally (H) polarized particles tunnel with probability $\tau_{\parallel}$ between $B$ and $C$ sites and with probability $\tau_{\perp}$ between $B$ and $A$ sites, since the projection of the polarization is longitudinal and transverse to the tunneling directions respectively [see Fig. \ref{fig_polarisation}(c)]. This means that the H-polarized eigenmodes of the $S$ flat band must have a higher population on $A$ sites, to compensate for the lower tunneling probability (Eq. \ref{tunnelling_relation}). The opposite holds for vertically (V) polarized eigenmodes, which will be characterized by a higher population on $C$ sites.

Finally, in our case, the quasi-resonant pump populates both linear combinations of H-polarized eigenmodes and linear combinations of V-polarized eigenmodes of the $S$ band. This, combined with the different populations of differently-polarized particles results in a nonzero degree of polarization on $A$ and $C$ sites. In order to confirm this explanation, we developed a TB model with polarization-dependent tunneling terms \cite{Note1}. By fitting the energy width of the $S$ band and the degree of polarization of the two sublattices we deduce the two hopping parameters to be: $\tau_{\parallel}=0.165$ meV and $\tau_{\perp}=0.145$ meV, in agreement with previous experimental results \cite{PhysRevX.5.011034}. With these values we obtain a degree of polarization of the order of 0.128, in excellent agreement with the experimental value of 0.13.

For the $P$ flat band the polarization pattern is qualitatively the same as that of the $S$ flat band albeit with a higher degree of polarization. It can be explained in the same way. In this case however one also has to consider that the $P$ state of each pillar is four-fold degenerate since, in addition to the polarization degeneracy, one also has the mode degeneracy of the $P_{x}$ and $P_{y}$ orbitals. It can be seen in Figs. \ref{fig1}(e) and \ref{intensity_dependence}(j) that for the $P$ flat band the emission predominantly has two lobes aligned along the $x$ direction on $A$ sites, corresponding to $P_{x}$ orbitals, while on the $C$ sites the $P_{y}$ orbitals dominate. We estimate the ratio of the orbital populations to be about 6:1, corresponding to $|\psi_{P_{x}}|^2/|\psi_{P_{y}}|^2$ on $A$ sites and $|\psi_{P_{y}}|^2/|\psi_{P_{x}}|^2$ on $C$ sites \cite{Note1}. As before, one needs to explain the suppression of emission from $B$ sites, since only orbitals with the same symmetry and polarization may destructively interfere. Similarly to the previously described polarization-dependent tunneling probability, one can observe that particles in $P_{x}$ orbitals tunnel more easily along the $x$ direction than along the $y$ direction, and therefore more particles in the $P_{x}$ orbital of $A$ sites are needed to satisfy Eq. (\ref{tunnelling_relation}), and the opposite holds for $P_{y}$ orbitals.

To confirm this we also developed a TB model for the four degenerate $P$ orbitals with tunneling amplitudes that depend on the polarization and on the alignment of the mode with respect to the hopping direction \cite{Note1}. As shown in Fig. \ref{fig_polarisation}(d) now we have four tunneling parameters: $\tau_{\parallel}^a$, $\tau_{\perp}^a$, $\tau_{\parallel}^t$, and $\tau_{\perp}^t$, where ($a$) and ($t$) indicate if the hopping is for $P$ orbitals aligned or transverse to the propagation direction, and ($\parallel$) and ($\perp$) indicate, as before, if the polarization is parallel or perpendicular to the hopping direction. Similarly to the $S$ band also here all four modes of each pillar will be populated by the quasi-resonant pump but on $A$ sites the population of $P_x$ H-polarized particles will be the highest, since their probability to tunnel to $B$ sites ($\tau_{\perp}^t$) is the lowest. Conversely, on $C$ sites, the population of $P_y$ V-polarized particles will be the highest. This is exactly what is observed in Fig. \ref{fig_polarisation}(b). By fitting the TB band structure to the experimentally observed $P$ band and degree of linear polarization the following hopping parameters can be obtained: $\tau_{\parallel}^a=0.375$, $\tau_{\perp}^a=0.125$, $\tau_{\parallel}^t=0.100$, and $\tau_{\perp}^t=0.033$ meV. Note that for the $P$ band the difference between the hopping with $\perp$ and $\parallel$ polarization is bigger than in the $S$ band case. This can be ascribed to the fact the $P$ flat band consists of harmonics with higher $k$ values, where polarization-dependent tunneling is expected to be enhanced \cite{Solnyshkov2016920}. With the values for the hopping parameters above we obtain a degree of polarization of the order of 0.42 and ratio between the populations of $P_x$ and $P_y$ orbitals on A sites of 4.1 (the inverse applies to $C$ sites), in good agreement with the experimental values of 0.5 and 6 \cite{Note1}. It should be noted that the tunneling arguments presented here apply equally in the single-particle regime, and as such polarization patterns are observed in the $S$ and $P$ flat-band emission below threshold. The polarization degree of the $P$ flat band was considerably lower ($\sim$0.2) than in the condensate regime however, probably due to contribution of the emission of the dispersive band at the same energy [cf. Fig. \ref{intensity_dependence}(g)].

In summary, we have studied the properties of a two-dimensional Lieb lattice for exciton-polaritons, demonstrating bosonic condensation into two separate flat bands formed from $S$ and $P$ orbitals, in addition to the negative effective mass states at the maxima of the $S$ anti-bonding band. 
We have also revealed distinctive emission patterns formed by the symmetric $S$ and asymmetric $P$ orbitals, which show pseudospin texture arising from spin-orbit coupling given by polarization-dependent tunneling between pillars. Our work shows the potential for engineering versatile lattice Hamiltonians for polaritons, highlighting the ease with which spin-orbit coupling terms and population of higher orbitals can be implemented, which presents a significant advantage of this system. Furthermore, the observation of flat-band condensate fragmentation demonstrates the effect of many-body interactions in the presence of quenched kinetic energy. An intriguing future prospect is studying quantum fluctuations as in recent polariton works \cite{PhysRevLett.118.247402,PhysRevX.7.031033,1707.01837}, in lattice environments where novel driven-dissipative phase transitions are expected \cite{PhysRevA.96.023839}.   

Currently the strength of polariton-polariton interactions in a single lattice site (few $\mu$eV) \cite{Walker2015} is comparable to or less than the polariton decay rate. However, the ratio of these two quantities may be further enhanced via polariton Feshbach resonances \cite{Takemura2014} or recently developed high-Q open-access microcavities with strong lateral confinement \cite{PhysRevLett.115.246401}. This would open the way to strongly correlated regimes described by driven-dissipative Bose-Hubbard models in polaritonic lattices \cite{Rota2017}. Such regimes are not accessible in weakly-coupled photonic systems.

The work was supported by EPSRC Grant EP/N031776/1, ERC Advanced Grant EXCIPOL No.
320570 and Megagrant N 14.Y26.31.0015. 
D.R.G. and I.V.I. acknowledge support from the project N 3.8884.2017 of the Ministry of Education and Science of Russian Federation. I.A.S. acknowledges support form the project N 3.2614.2017 of the Ministry of Education and Science of Russian Federation and Rannis project 163082-051.

\bibliographystyle{ieeetr}
\bibliography{manual,manual2}   

\begin{thebibliography}{10}

\bibitem{PhysRevA.82.053618}
S.~Zhang, H.-h. Hung, and C.~Wu, ``Proposed realization of itinerant
  ferromagnetism in optical lattices,'' {\em Phys. Rev. A}, vol.~82, p.~053618,
  Nov 2010.

\bibitem{PhysRevLett.99.070401}
C.~Wu, D.~Bergman, L.~Balents, and S.~Das~Sarma, ``Flat bands and wigner
  crystallization in the honeycomb optical lattice,'' {\em Phys. Rev. Lett.},
  vol.~99, p.~070401, Aug 2007.

\bibitem{PhysRevLett.107.146803}
Y.-F. Wang, Z.-C. Gu, C.-D. Gong, and D.~N. Sheng, ``Fractional quantum hall
  effect of hard-core bosons in topological flat bands,'' {\em Phys. Rev.
  Lett.}, vol.~107, p.~146803, Sep 2011.

\bibitem{PhysRevLett.62.1201}
E.~H. Lieb, ``Two theorems on the hubbard model,'' {\em Phys. Rev. Lett.},
  vol.~62, pp.~1201--1204, Mar 1989.

\bibitem{Keimer}
B.~Keimer, S.~A. Kivelson, M.~R. Norman, S.~Uchida, and J.~Zaanen, ``From
  quantum matter to high-temperature superconductivity in copper oxides,'' {\em
  Nature}, vol.~518, pp.~179--186, Feb 2015.
\newblock Review.

\bibitem{PhysRevB.82.085310}
C.~Weeks and M.~Franz, ``Topological insulators on the lieb and perovskite
  lattices,'' {\em Phys. Rev. B}, vol.~82, p.~085310, Aug 2010.

\bibitem{PhysRevA.83.063601}
N.~Goldman, D.~F. Urban, and D.~Bercioux, ``Topological phases for fermionic
  cold atoms on the lieb lattice,'' {\em Phys. Rev. A}, vol.~83, p.~063601, Jun
  2011.

\bibitem{PhysRevB.86.195129}
W.~Beugeling, J.~C. Everts, and C.~Morais~Smith, ``Topological phase
  transitions driven by next-nearest-neighbor hopping in two-dimensional
  lattices,'' {\em Phys. Rev. B}, vol.~86, p.~195129, Nov 2012.

\bibitem{doi:10.1142/S021797921330017X}
E.~J. Bergholtz and Z.~Liu, ``Topological flat band models and fractional chern
  insulators,'' {\em International Journal of Modern Physics B}, vol.~27,
  no.~24, p.~1330017, 2013.

\bibitem{1367-2630-17-5-055016}
W.-F. Tsai, C.~Fang, H.~Yao, and J.~Hu, ``Interaction-driven topological and
  nematic phases on the lieb lattice,'' {\em New Journal of Physics}, vol.~17,
  no.~5, p.~055016, 2015.

\bibitem{2015arXiv151007239K}
I.~N. {Karnaukhov} and I.~O. {Slieptsov}, ``{The Chern states on the honeycomb
  and Lieb lattices},'' {\em ArXiv e-prints}, Oct. 2015.

\bibitem{PhysRevB.92.235106}
G.~Palumbo and K.~Meichanetzidis, ``Two-dimensional chern semimetals on the
  lieb lattice,'' {\em Phys. Rev. B}, vol.~92, p.~235106, Dec 2015.

\bibitem{PhysRevA.93.043611}
A.~Dauphin, M.~M\"uller, and M.~A. Martin-Delgado, ``Quantum simulation of a
  topological mott insulator with rydberg atoms in a lieb lattice,'' {\em Phys.
  Rev. A}, vol.~93, p.~043611, Apr 2016.

\bibitem{1674-1056-25-6-067204}
R.~Chen and B.~Zhou, ``Finite size effects on the helical edge states on the
  lieb lattice,'' {\em Chinese Physics B}, vol.~25, no.~6, p.~067204, 2016.

\bibitem{WangR2016}
R.~Wang, Q.~Qiao, B.~Wang, X.-H. Ding, and Y.-F. Zhang, ``The topological
  quantum phase transitions in lieb lattice driven by the rashba soc and
  exchange field,'' {\em The European Physical Journal B}, vol.~89, no.~9,
  p.~192, 2016.

\bibitem{PhysRevLett.117.163001}
M.~Di~Liberto, A.~Hemmerich, and C.~Morais~Smith, ``Topological varma
  superfluid in optical lattices,'' {\em Phys. Rev. Lett.}, vol.~117,
  p.~163001, Oct 2016.

\bibitem{1367-2630-16-6-063061}
D.~Guzmán-Silva, C.~Mejía-Cortés, M.~A. Bandres, M.~C. Rechtsman,
  S.~Weimann, S.~Nolte, M.~Segev, A.~Szameit, and R.~A. Vicencio,
  ``Experimental observation of bulk and edge transport in photonic lieb
  lattices,'' {\em New Journal of Physics}, vol.~16, no.~6, p.~063061, 2014.

\bibitem{PhysRevLett.114.245504}
S.~Mukherjee, A.~Spracklen, D.~Choudhury, N.~Goldman, P.~\"Ohberg,
  E.~Andersson, and R.~R. Thomson, ``Observation of a localized flat-band state
  in a photonic lieb lattice,'' {\em Phys. Rev. Lett.}, vol.~114, p.~245504,
  Jun 2015.

\bibitem{PhysRevLett.116.183902}
F.~Diebel, D.~Leykam, S.~Kroesen, C.~Denz, and A.~S. Desyatnikov, ``Conical
  diffraction and composite lieb bosons in photonic lattices,'' {\em Phys. Rev.
  Lett.}, vol.~116, p.~183902, May 2016.

\bibitem{2053-1583-4-2-025008}
C.~Poli, H.~Schomerus, M.~Bellec, U.~Kuhl, and F.~Mortessagne, ``Partial chiral
  symmetry-breaking as a route to spectrally isolated topological defect states
  in two-dimensional artificial materials,'' {\em 2D Materials}, vol.~4, no.~2,
  p.~025008, 2017.

\bibitem{Taiee1500854}
S.~Taie, H.~Ozawa, T.~Ichinose, T.~Nishio, S.~Nakajima, and Y.~Takahashi,
  ``Coherent driving and freezing of bosonic matter wave in an optical lieb
  lattice,'' {\em Science Advances}, vol.~1, no.~10, 2015.

\bibitem{PhysRevA.93.043847}
D.~L\'opez-Gonz\'alez and M.~I. Molina, ``Linear and nonlinear compact modes in
  quasi-one-dimensional flatband systems,'' {\em Phys. Rev. A}, vol.~93,
  p.~043847, Apr 2016.

\bibitem{PhysRevB.94.144302}
G.~Gligori\ifmmode~\acute{c}\else \'{c}\fi{}, A.~Maluckov,
  L.~Had\ifmmode~\check{z}\else \v{z}\fi{}ievski, S.~Flach, and B.~A. Malomed,
  ``Nonlinear localized flat-band modes with spin-orbit coupling,'' {\em Phys.
  Rev. B}, vol.~94, p.~144302, Oct 2016.

\bibitem{Amo2016934}
A.~Amo and J.~Bloch, ``Exciton-polaritons in lattices: A non-linear photonic
  simulator,'' {\em Comptes Rendus Physique}, vol.~17, no.~8, pp.~934 -- 945,
  2016.
\newblock Polariton physics / Physique des polaritons.

\bibitem{KasprzakJ2006}
J.~Kasprzak, M.~Richard, S.~Kundermann, A.~Baas, P.~Jeambrun, J.~M.~J. Keeling,
  F.~M. Marchetti, M.~H. Szymanska, R.~Andre, J.~L. Staehli, V.~Savona, P.~B.
  Littlewood, B.~Deveaud, and L.~S. Dang, ``Bose-einstein condensation of
  exciton polaritons,'' {\em Nature}, vol.~443, pp.~409--414, Sep 2006.

\bibitem{Balili1007}
R.~Balili, V.~Hartwell, D.~Snoke, L.~Pfeiffer, and K.~West, ``Bose-einstein
  condensation of microcavity polaritons in a trap,'' {\em Science}, vol.~316,
  no.~5827, pp.~1007--1010, 2007.

\bibitem{PhysRevLett.101.067404}
A.~P.~D. Love, D.~N. Krizhanovskii, D.~M. Whittaker, R.~Bouchekioua,
  D.~Sanvitto, S.~A. Rizeiqi, R.~Bradley, M.~S. Skolnick, P.~R. Eastham,
  R.~Andr\'e, and L.~S. Dang, ``Intrinsic decoherence mechanisms in the
  microcavity polariton condensate,'' {\em Phys. Rev. Lett.}, vol.~101,
  p.~067404, Aug 2008.

\bibitem{PhysRevB.80.045317}
D.~N. Krizhanovskii, K.~G. Lagoudakis, M.~Wouters, B.~Pietka, R.~A. Bradley,
  K.~Guda, D.~M. Whittaker, M.~S. Skolnick, B.~Deveaud-Pl\'edran, M.~Richard,
  R.~Andr\'e, and L.~S. Dang, ``Coexisting nonequilibrium condensates with
  long-range spatial coherence in semiconductor microcavities,'' {\em Phys.
  Rev. B}, vol.~80, p.~045317, Jul 2009.

\bibitem{Walker2015}
P.~M. Walker, L.~Tinkler, D.~V. Skryabin, A.~Yulin, B.~Royall, I.~Farrer, D.~A.
  Ritchie, M.~S. Skolnick, and D.~N. Krizhanovskii, ``Ultra-low-power hybrid
  light-matter solitons,'' {\em Nature Communications}, vol.~6, pp.~8317 EP --,
  Sep 2015.
\newblock Article.

\bibitem{SICH2016908}
M.~Sich, D.~V. Skryabin, and D.~N. Krizhanovskii, ``Soliton physics with
  semiconductor exciton–polaritons in confined systems,'' {\em Comptes Rendus
  Physique}, vol.~17, no.~8, pp.~908 -- 919, 2016.
\newblock Polariton physics / Physique des polaritons.

\bibitem{proukakis_snoke_littlewood_2017}
{\em Universal Themes of Bose-Einstein Condensation}.
\newblock Cambridge University Press, 2017.

\bibitem{PhysRevLett.118.247402}
S.~R.~K. Rodriguez, W.~Casteels, F.~Storme, N.~Carlon~Zambon, I.~Sagnes,
  L.~Le~Gratiet, E.~Galopin, A.~Lema\^{\i}tre, A.~Amo, C.~Ciuti, and J.~Bloch,
  ``Probing a dissipative phase transition via dynamical optical hysteresis,''
  {\em Phys. Rev. Lett.}, vol.~118, p.~247402, Jun 2017.

\bibitem{PhysRevX.7.031033}
C.~E. Whittaker, B.~Dzurnak, O.~A. Egorov, G.~Buonaiuto, P.~M. Walker,
  E.~Cancellieri, D.~M. Whittaker, E.~Clarke, S.~S. Gavrilov, M.~S. Skolnick,
  and D.~N. Krizhanovskii, ``Polariton pattern formation and photon statistics
  of the associated emission,'' {\em Phys. Rev. X}, vol.~7, p.~031033, Aug
  2017.

\bibitem{1707.01837}
T.~Fink, A.~Schade, S.~Höfling, C.~Schneider, and A.~İmamoğlu, ``Signatures
  of a dissipative phase transition in photon correlation measurements,'' 2017.

\bibitem{KimNY2011}
N.~Y. Kim, K.~Kusudo, C.~Wu, N.~Masumoto, A.~Loffler, S.~Hofling, N.~Kumada,
  L.~Worschech, A.~Forchel, and Y.~Yamamoto, ``Dynamical d-wave condensation of
  exciton-polaritons in a two-dimensional square-lattice potential,'' {\em Nat
  Phys}, vol.~7, pp.~681--686, Sep 2011.

\bibitem{PhysRevLett.112.116402}
T.~Jacqmin, I.~Carusotto, I.~Sagnes, M.~Abbarchi, D.~D. Solnyshkov,
  G.~Malpuech, E.~Galopin, A.~Lema\^{\i}tre, J.~Bloch, and A.~Amo, ``Direct
  observation of dirac cones and a flatband in a honeycomb lattice for
  polaritons,'' {\em Phys. Rev. Lett.}, vol.~112, p.~116402, Mar 2014.

\bibitem{1367-2630-17-2-023001}
K.~Winkler, J.~Fischer, A.~Schade, M.~Amthor, R.~Dall, J.~Geßler,
  M.~Emmerling, E.~A. Ostrovskaya, M.~Kamp, C.~Schneider, and S.~Höfling, ``A
  polariton condensate in a photonic crystal potential landscape,'' {\em New
  Journal of Physics}, vol.~17, no.~2, p.~023001, 2015.

\bibitem{PhysRevLett.105.116402}
E.~A. Cerda-M\'endez, D.~N. Krizhanovskii, M.~Wouters, R.~Bradley, K.~Biermann,
  K.~Guda, R.~Hey, P.~V. Santos, D.~Sarkar, and M.~S. Skolnick, ``Polariton
  condensation in dynamic acoustic lattices,'' {\em Phys. Rev. Lett.},
  vol.~105, p.~116402, Sep 2010.

\bibitem{PhysRevB.87.155423}
D.~N. Krizhanovskii, E.~A. Cerda-M\'endez, S.~Gavrilov, D.~Sarkar, K.~Guda,
  R.~Bradley, P.~V. Santos, R.~Hey, K.~Biermann, M.~Sich, F.~Fras, and M.~S.
  Skolnick, ``Effect of polariton-polariton interactions on the excitation
  spectrum of a nonequilibrium condensate in a periodic potential,'' {\em Phys.
  Rev. B}, vol.~87, p.~155423, Apr 2013.

\bibitem{PhysRevB.86.100301}
E.~A. Cerda-M\'endez, D.~N. Krizhanovskii, K.~Biermann, R.~Hey, M.~S. Skolnick,
  and P.~V. Santos, ``Dynamic exciton-polariton macroscopic coherent phases in
  a tunable dot lattice,'' {\em Phys. Rev. B}, vol.~86, p.~100301, Sep 2012.

\bibitem{TosiG2012}
G.~Tosi, G.~Christmann, N.~G. Berloff, P.~Tsotsis, T.~Gao, Z.~Hatzopoulos,
  P.~G. Savvidis, and J.~J. Baumberg, ``Sculpting oscillators with light within
  a nonlinear quantum fluid,'' {\em Nat Phys}, vol.~8, pp.~190--194, Mar 2012.

\bibitem{PhysRevLett.118.107403}
M.~Mili\ifmmode \acute{c}\else \'{c}\fi{}evi\ifmmode~\acute{c}\else \'{c}\fi{},
  T.~Ozawa, G.~Montambaux, I.~Carusotto, E.~Galopin, A.~Lema\^{\i}tre,
  L.~Le~Gratiet, I.~Sagnes, J.~Bloch, and A.~Amo, ``Orbital edge states in a
  photonic honeycomb lattice,'' {\em Phys. Rev. Lett.}, vol.~118, p.~107403,
  Mar 2017.

\bibitem{PhysRevX.5.011034}
V.~G. Sala, D.~D. Solnyshkov, I.~Carusotto, T.~Jacqmin, A.~Lema\^{\i}tre,
  H.~Ter\ifmmode~\mbox{\c{c}}\else \c{c}\fi{}as, A.~Nalitov, M.~Abbarchi,
  E.~Galopin, I.~Sagnes, J.~Bloch, G.~Malpuech, and A.~Amo, ``Spin-orbit
  coupling for photons and polaritons in microstructures,'' {\em Phys. Rev. X},
  vol.~5, p.~011034, Mar 2015.

\bibitem{PhysRevLett.115.246401}
S.~Dufferwiel, F.~Li, E.~Cancellieri, L.~Giriunas, A.~A.~P. Trichet, D.~M.
  Whittaker, P.~M. Walker, F.~Fras, E.~Clarke, J.~M. Smith, M.~S. Skolnick, and
  D.~N. Krizhanovskii, ``Spin textures of exciton-polaritons in a tunable
  microcavity with large te-tm splitting,'' {\em Phys. Rev. Lett.}, vol.~115,
  p.~246401, Dec 2015.

\bibitem{Solnyshkov2016920}
D.~Solnyshkov and G.~Malpuech, ``Chirality in photonic systems,'' {\em Comptes
  Rendus Physique}, vol.~17, no.~8, pp.~920 -- 933, 2016.
\newblock Polariton physics / Physique des polaritons.

\bibitem{0268-1242-25-1-013001}
I.~A. Shelykh, A.~V. Kavokin, Y.~G. Rubo, T.~C.~H. Liew, and G.~Malpuech,
  ``Polariton polarization-sensitive phenomena in planar semiconductor
  microcavities,'' {\em Semiconductor Science and Technology}, vol.~25, no.~1,
  p.~013001, 2010.

\bibitem{PhysRevLett.116.066402}
F.~Baboux, L.~Ge, T.~Jacqmin, M.~Biondi, E.~Galopin, A.~Lema\^{\i}tre,
  L.~Le~Gratiet, I.~Sagnes, S.~Schmidt, H.~E. T\"ureci, A.~Amo, and J.~Bloch,
  ``Bosonic condensation and disorder-induced localization in a flat band,''
  {\em Phys. Rev. Lett.}, vol.~116, p.~066402, Feb 2016.

\bibitem{PhysRevB.34.5208}
B.~Sutherland, ``Localization of electronic wave functions due to local
  topology,'' {\em Phys. Rev. B}, vol.~34, pp.~5208--5211, Oct 1986.

\bibitem{Note1}
See supplementary material at [url] for further details about the sample,
  experiment and theoretical models, which includes Ref. \cite
  {PhysRevB.76.201305}.

\bibitem{PhysRevB.51.13614}
E.~L. Shirley, L.~J. Terminello, A.~Santoni, and F.~J. Himpsel,
  ``Brillouin-zone-selection effects in graphite photoelectron angular
  distributions,'' {\em Phys. Rev. B}, vol.~51, pp.~13614--13622, May 1995.

\bibitem{PhysRevB.87.125428}
M.~Ni\ifmmode \mbox{\c{t}}\else \c{t}\fi{}\ifmmode~\u{a}\else \u{a}\fi{},
  B.~Ostahie, and A.~Aldea, ``Spectral and transport properties of the
  two-dimensional lieb lattice,'' {\em Phys. Rev. B}, vol.~87, p.~125428, Mar
  2013.

\bibitem{PhysRevLett.97.097402}
D.~N. Krizhanovskii, D.~Sanvitto, A.~P.~D. Love, M.~S. Skolnick, D.~M.
  Whittaker, and J.~S. Roberts, ``Dominant effect of polariton-polariton
  interactions on the coherence of the microcavity optical parametric
  oscillator,'' {\em Phys. Rev. Lett.}, vol.~97, p.~097402, Aug 2006.

\bibitem{TaneseD2013}
D.~Tanese, H.~Flayac, D.~Solnyshkov, A.~Amo, A.~Lema{\^i}tre, E.~Galopin,
  R.~Braive, P.~Senellart, I.~Sagnes, G.~Malpuech, and J.~Bloch, ``Polariton
  condensation in solitonic gap states in a one-dimensional periodic
  potential,'' {\em Nature Communications}, vol.~4, pp.~1749 EP --, Apr 2013.
\newblock Article.

\bibitem{PhysRevLett.111.146401}
E.~A. Cerda-M\'endez, D.~Sarkar, D.~N. Krizhanovskii, S.~S. Gavrilov,
  K.~Biermann, M.~S. Skolnick, and P.~V. Santos, ``Exciton-polariton gap
  solitons in two-dimensional lattices,'' {\em Phys. Rev. Lett.}, vol.~111,
  p.~146401, Oct 2013.

\bibitem{PhysRevB.82.184502}
S.~D. Huber and E.~Altman, ``Bose condensation in flat bands,'' {\em Phys. Rev.
  B}, vol.~82, p.~184502, Nov 2010.

\bibitem{PhysRevLett.114.026803}
A.~V. Nalitov, G.~Malpuech, H.~Ter\ifmmode~\mbox{\c{c}}\else \c{c}\fi{}as, and
  D.~D. Solnyshkov, ``Spin-orbit coupling and the optical spin hall effect in
  photonic graphene,'' {\em Phys. Rev. Lett.}, vol.~114, p.~026803, Jan 2015.

\bibitem{PhysRevA.96.023839}
A.~Biella, F.~Storme, J.~Lebreuilly, D.~Rossini, R.~Fazio, I.~Carusotto, and
  C.~Ciuti, ``Phase diagram of incoherently driven strongly correlated photonic
  lattices,'' {\em Phys. Rev. A}, vol.~96, p.~023839, Aug 2017.

\bibitem{Takemura2014}
N.~Takemura, S.~Trebaol, M.~Wouters, M.~T. Portella-Oberli, and B.~Deveaud,
  ``Polaritonic feshbach resonance,'' {\em Nature Physics}, vol.~10, pp.~500 EP
  --, Jun 2014.

\bibitem{Rota2017}
R.~Rota, W.~Casteels, and C.~Ciuti, ``On the robustness of strongly correlated
  multi-photon states in frustrated driven-dissipative cavity lattices,'' {\em
  The European Physical Journal Special Topics}, vol.~226, pp.~2805--2814, Jul
  2017.

\bibitem{PhysRevB.76.201305}
D.~Bajoni, P.~Senellart, A.~Lema\^{\i}tre, and J.~Bloch, ``Photon lasing in
  $\mathrm{GaAs}$ microcavity: Similarities with a polariton condensate,'' {\em
  Phys. Rev. B}, vol.~76, p.~201305, Nov 2007.

\bibitem{Note2}
We use a standard bootstrap method and calculate the Pearson correlation
  coefficient $r$ between the normalized intensity and $\Delta $E for both the
  $S$ and $P$ flat bands, taking 1000 resamples of the data. This yields values
  of $r = 0.4139$ and $r = 0.7672$ for the $S$ and $P$ flat bands respectively.

\end{thebibliography}

\pagebreak
\widetext
\begin{center}
\textbf{\large Supplementary material}
\end{center}
\setcounter{equation}{0}
\setcounter{figure}{0}
\setcounter{table}{0}
\setcounter{page}{1}
\makeatletter
\renewcommand{\theequation}{S\arabic{equation}}
\renewcommand{\thefigure}{S\arabic{figure}}

\section{Sample details}

\begin{figure}[!htb]
\center
\includegraphics[scale=0.42]{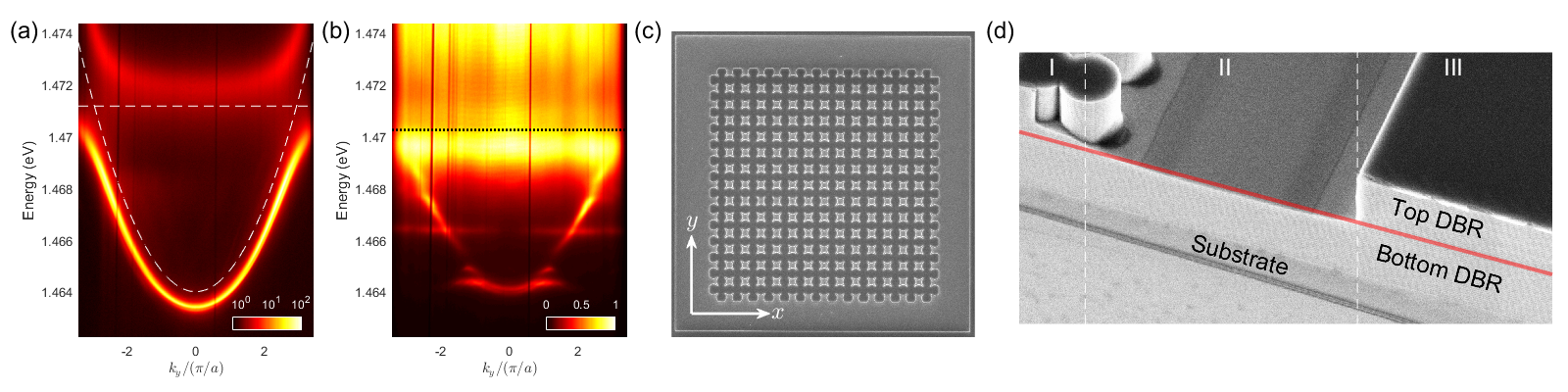}
\center
\vspace{-0.4cm}
\caption{\label{fig1} Angle-resolved PL spectra from the sample on a planar unetched (a) and lattice etched (b) region. (c) Arial SEM image of the whole lattice region. (d) Angled SEM image showing a cleaved piece of wafer with both planar and lattice regions.}
\label{lattice}
\end{figure}

Here we provide further details about our sample. Our 2D Lieb lattice was fabricated by processing (through a combination of electron beam lithography and plasma dry etching) a $\lambda/2$ GaAs microcavity with 23 (27) GaAs/Al$_{0.85}$Ga$_{0.15}$As pairs in the top (bottom) distributed Bragg reflector, and 3 In$_{0.04}$Ga$_{0.96}$As quantum wells placed at the antinode of the cavity field. In Fig. \ref{lattice}(a) we show an angle-resolved photoluminescence (PL) spectrum from the unetched planar region of cavity directly next to the etched lattice. The dashed white lines show the energies of the uncoupled cavity photon and exciton modes. From fitted curves to the polariton branches we estimate a Rabi splitting of around 4.7 meV, a Q factor of 14,000, a linewidth of around 100 $\mu$eV and a photon-exciton detuning of -7.2 meV. In Fig. \ref{lattice}(b) we show the energy spectrum from the etched lattice region of the sample studied in the main text. At high energies approaching the exciton we see a broad continuum band of higher energy pillar modes. The dotted line marks the energy of the pulsed laser used for quasi-resonant excitation in the main text. A view of the entire lattice can be seen in Fig. \ref{lattice}(c) which shows a scanning electron microscope (SEM) image. Excluding the pillars on the left and bottom edges, the lattice covers $14\times14$ unit cells. Fig. \ref{lattice}(d) shows an angled SEM image of an etched lattice which has been cleaved through the center to reveal the etch depth. On the left (region I) we see the edge of a lattice and on the right (region III) an unetched region of the wafer, between which is an etched region with no pillars (region II). The red line indicates the position of the active layer. It can be seen that the number of distributed Bragg reflector (DBR) pairs left after etching is inhomogeneous along this cut of the wafer, such that in region II the wafer is etched down through the active layer, whilst in region I a few layers of the top DBR remain intact. We estimate a top DBR thickness of 4-6 pairs in the lattice region, which we use in our Schr\"{o}dinger equation model later in the text.

\section{Experimental details}

In this section further details about the experiments reported in the main text are provided. Our sample is mounted in a continuous flow cryostat held at 8K where rear and front windows with large angular access allow us to work in both reflection and transmission geometry. For non-resonant excitation measurements, we use a continuous-wave diode laser operating at 685 nm with a spot size of 15 $\mu$m, collecting the reflected photoluminescence (PL) through the same N = 0.42 microscope objective (MO) which irradiates the sample. 

\begin{figure}[!htb]
\center
\includegraphics[scale=0.4]{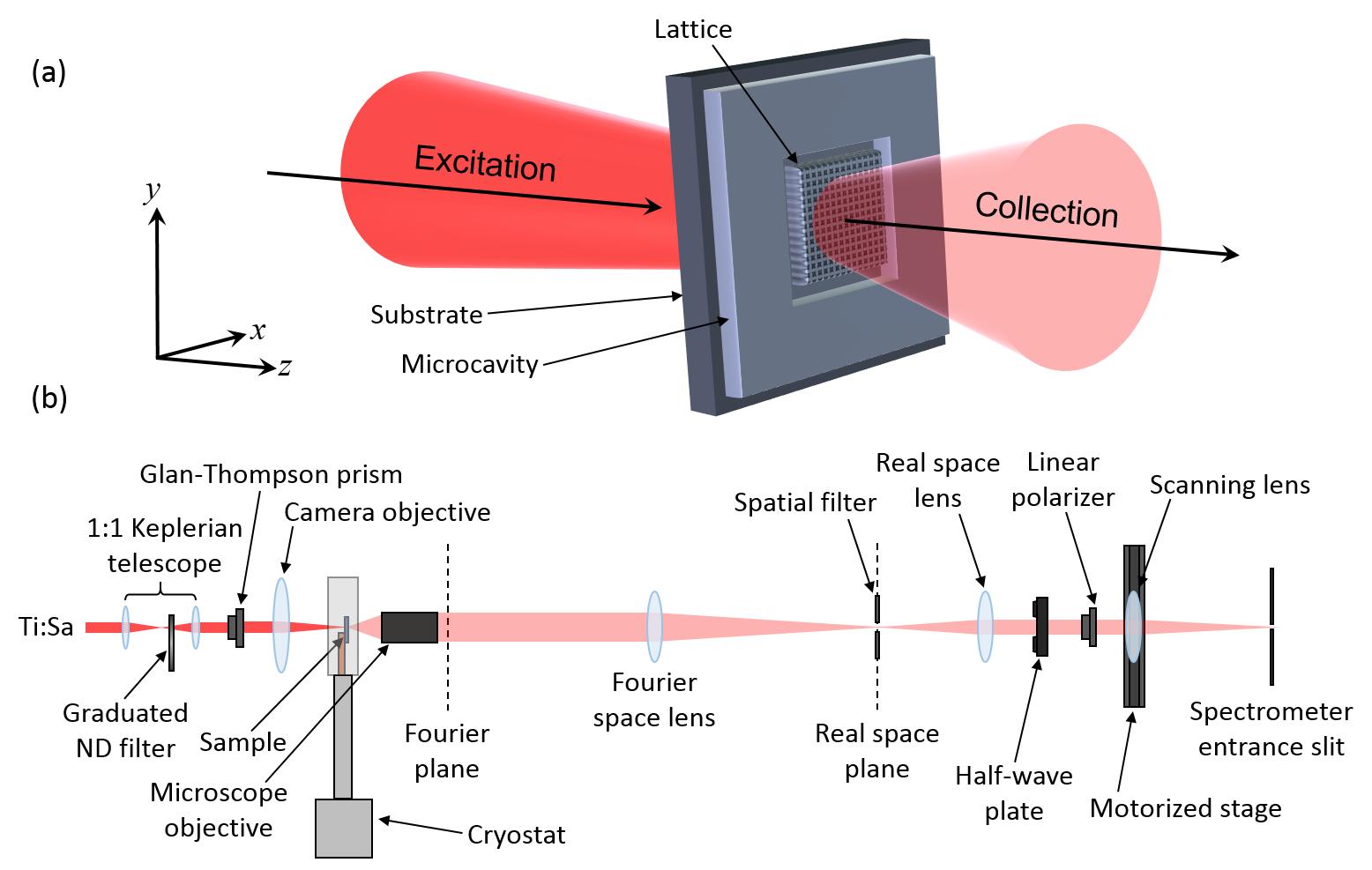}
\center
\vspace{-0.4cm}
\caption{\label{fig1} Schematic of the experimental setup. (a) Close-up of the substrate-side excitation scheme. (b) Arial view of the excitation and collection paths.}
\label{schema}
\end{figure}

For quasi-resonant excitation, we pump the substrate side of the sample with horizontally-polarized 100 ps pulses generated by a mode-locked Ti:Sapphire laser with an 80 MHz repetition rate, using a camera objective with a 5 cm focal length to focus on the sample. A schematic of the experimental setup can be seen in Fig. \ref{schema}. In the excitation path, a graduated circular neutral density filter is mounted on a motorized rotation stage placed slightly displaced from the laser beam waist inside a 1:1 Keplerian telescope. A Glan-Thompson polarizer horizontally polarizes the laser pulses, which are then focused onto the sample using a camera objective. The emitted photoluminescence is collected by a microscope objective (MO), before passing through a confocal 30 cm lens and a subsequent 10 cm lens on a flip mount. In the confocal plane between the second and third lens, two orthogonal adjustable slits allow for real space filtering from the magnified virtual image plane. The final lens in the setup is a 15 cm scanning lens. For Fourier space measurements, the third lens is flipped out of the optical path to project an image of the MO back focal plane onto the spectrometer entrance slit. For real space measurements, the third lens is flipped into position, forming near-field images with a magnification of $\times$45. For polarization measurements we cross-polarized a linear polarizer to the excitation laser, and found the angles for the fast and slow axis of a half wave plate mounted on a motorized rotation stage to resolve the emission in the horizontal-vertical basis. 

\section{Tight-binding model}

In order to explain the band structure and the polarization pattern observed in our Lieb lattice we develop a tight-binding (TB) model, for both the $S$ and $P$ bands, with polarization-dependent hopping amplitudes. For our model we only consider nearest-neighbor terms, since the contribution from higher order terms is negligible \cite{PhysRevX.5.011034}. As the $P$ bands are formed from coupling of the first excited pillar mode, which is a four-fold degenerate orbital (doubly-degenerate due to the dipole structure and doubly-degenerate due to the polarization), we also include orientation-dependent hopping probabilities, such that the coupling depends on whether the lobes are oriented longitudinally or transversely to the tunneling direction \cite{PhysRevLett.99.070401}. Once the kernel matrices for the two TB models are found, it is possible to calculate the polarization-resolved mode occupation on each pillar, and calculate the corresponding linear polarization degree.

\subsection{$S$ band}
By defining $a_{H,m,n}$ and $a_{V,m,n}$ as the annihilation operators in the $S$ orbital modes of the $A$ sublattice pillars with linear polarization along the horizontal and vertical directions respectively (and similarly for the $B$ and $C$ sites), the Hamiltonian for the $S$ band can be written as:

\begin{eqnarray*}
H_{Lieb}^{S}&=&
-\sum_{m,n=-\infty}^{\infty}b_{H,m,n}^{\dagger}[\tau_{\perp}(a_{H,m,n}+a_{H,m-1,n})+\tau_{\parallel}(c_{H,m,n}+c_{H,m,n-1})]+
\\&&
\hspace{1.7cm}b_{V,m,n}^{\dagger}[\tau_{\parallel}(a_{V,m,n}+a_{V,m-1,n})+\tau_{\perp}(c_{V,m,n}+c_{V,m,n-1})]+h.c.,
\end{eqnarray*}

\noindent
where the hopping probabilities are $\tau_{\parallel}$, $\tau_{\perp}$ when the polarization is aligned along (parallel) or sideways to (perpendicular) the hopping direction. The on-site $S$ orbital energies are all equal and set to zero. Introducing the Fourier transform of the creation and annihilation operators as:

\begin{equation*}
a_{H,p,q}=\frac{1}{N}\sum_{m,n=-\infty}^{\infty}a_{H,m,n}e^{+i\alpha(k_pm+k_qn)},\hspace{1cm}a_{H,p,q}^{\dagger}=\frac{1}{N}\sum_{m,n=-\infty}^{\infty}a_{H,m,n}^{\dagger}e^{-i\alpha(k_pm+k_qn)},
\end{equation*}

\noindent
where $\alpha$ is the unit cell size of the TB model, and $k_p$ and $k_q$ are the $x$ and $y$ component of the wave-vector, respectively. Similarly it is possible to define the Fourier transform of the operators for the $B$ and $C$ sites and for the V polarization, it is possible to write the above Hamiltonian in $k$-space. This can be written in a compact form as:

\begin{equation*}
H_{Lieb}^{S}=-2\sum_{p,q=-\infty}^{\infty}\psi_{p,q}^{\dagger^T}
\begin{pmatrix}
M_{H,p,q}^S & 0 \\ 0 & M_{V,p,q}^S
\end{pmatrix}\psi_{p,q},
\end{equation*}

\noindent
with

\begin{equation*}
\psi_{p,q}^{\dagger^T}=(a^{\dagger}_{H,p,q} , b^{\dagger}_{H,p,q} , c^{\dagger}_{H,p,q} , a^{\dagger}_{V,p,q} , b^{\dagger}_{V,p,q} , c^{\dagger}_{V,p,q})
\end{equation*}

\noindent
and

\begin{equation*}
M_{H,p,q}^S=
\begin{pmatrix}
0 & \tau_{\perp}e^{+\frac{i\alpha k_q}{2}}\cos\left(\frac{k_q\alpha}{2}\right) & 0 \\
\tau_{\perp}^{\star}e^{-\frac{i\alpha k_q}{2}}\cos\left(\frac{k_q\alpha}{2}\right)  & 0 & \tau_{\parallel}e^{+\frac{i\alpha k_p}{2}}\cos\left(\frac{k_p\alpha}{2}\right) \\
0 & \tau_{\parallel}^{\star}e^{-\frac{i\alpha k_p}{2}}\cos\left(\frac{k_p\alpha}{2}\right) & 0 
\end{pmatrix},
\end{equation*}

\noindent
and

\begin{equation*}
M_{V,p,q}^S=
\begin{pmatrix}
0 & \tau_{\parallel}e^{+\frac{i\alpha k_q}{2}}\cos\left(\frac{k_q\alpha}{2}\right) & 0 \\
\tau_{\parallel}^{\star}e^{-\frac{i\alpha k_q}{2}}\cos\left(\frac{k_q\alpha}{2}\right)  & 0 & \tau_{\perp}e^{+\frac{i\alpha k_p}{2}}\cos\left(\frac{k_p\alpha}{2}\right) \\
0 & \tau_{\perp}^{\star}e^{-\frac{i\alpha k_p}{2}}\cos\left(\frac{k_p\alpha}{2}\right) & 0 
\end{pmatrix}.
\end{equation*}

The polarized eigenvector of the $S$ band can be easily written in terms of the function $E[x,y]=-2xe^{-\frac{i\alpha y}{2}}cos\left(\frac{y\alpha}{2}\right)$, where $x$ is the hopping probability and $y$ the direction of propagation:

\begin{eqnarray*}
e_1&=&\frac{1}{A_1}\left\{0,-\frac{E[\tau_{\perp},k_q]}{E[\tau_{\parallel},k_p]},1,0,0,0\right\}\\
e_2&=&\frac{1}{A_2}\left\{-\frac{\sqrt{|E[\tau_{\parallel},k_p]|^2+|E[\tau_{\perp},k_q]|^2}}{E[\tau_{\perp},k_q]^*},\frac{E[\tau_{\parallel},k_p]^*}{E[\tau_{\perp},k_q]^*},1,0,0,0\right\}\\
e_3&=&\frac{1}{A_3}\left\{+\frac{\sqrt{|E[\tau_{\parallel},k_p]|^2+|E[\tau_{\perp},k_q]|^2}}{E[\tau_{\perp},k_q]^*},\frac{E[\tau_{\parallel},k_p]^*}{E[\tau_{\perp},k_q]^*},1,0,0,0\right\}\\
e_4&=&\frac{1}{A_4}\left\{0,0,0,0,-\frac{E[\tau_{\parallel},k_q]}{E[\tau_{\perp},k_p]},1\right\}\\
e_5&=&\frac{1}{A_5}\left\{0,0,0,-\frac{\sqrt{|E[\tau_{\parallel},k_q]|^2+|E[\tau_{\perp},k_p]|^2}}{E[\tau_{\parallel},k_q]^*},\frac{E[\tau_{\perp},k_p]^*}{E[\tau_{\parallel},k_q]^*},1\right\}\\
e_6&=&\frac{1}{A_6}\left\{0,0,0,+\frac{\sqrt{|E[\tau_{\parallel},k_q]|^2+|E[\tau_{\perp},k_p]|^2}}{E[\tau_{\parallel},k_q]^*},\frac{E[\tau_{\perp},k_p]^*}{E[\tau_{\parallel},k_q]^*},1\right\},
\end{eqnarray*}

\noindent where the six constants $A_i$ are normalization constants. We see here that with these assumptions the H and V polarized modes are completely independent since the kernel matrix is a block matrix. Each block is basically the $3\times3$ kernel matrix for a single mode Lieb lattice. The main difference here is that the hopping probabilities change depending on the polarization and on the hopping direction.

\begin{figure}[!hb]
\center
\hspace*{-1cm}
\includegraphics[scale=0.4]{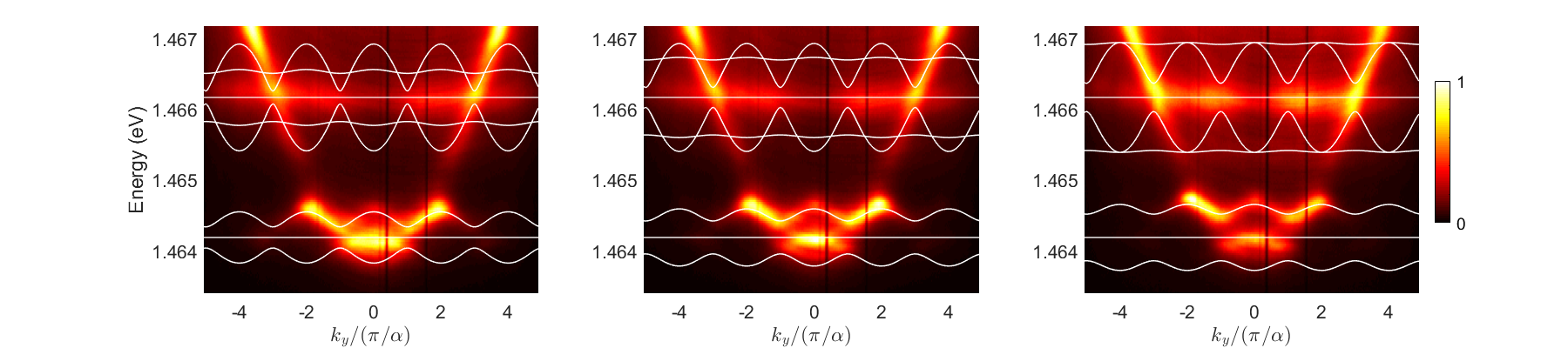}
\caption{\label{fig-both-bands}
Experimentally-measured energy-momentum relations at $k_x/(\pi/\alpha)$ = 1.3, 1.5 and 2, overlaid with curves from the TB model (not including polarization).}
\end{figure}

In order to reproduce and explain the experimental observations we first consider the unpolarized case (i.e. $\tau_{\perp}=\tau_{\parallel}=\tau$) and fit the $S$ part of the experimental band structure. The result of this fitting is shown in the theoretical curves overlapped to the experimental band in Fig. 1(b) of the main text and in Fig. \ref{fig-both-bands}. From this fitting we obtained a value of $\tau=0.165$ meV. Note that since we have considered identical values for the two polarizations, the theoretical lines overlapped to the $S$ band in Fig. 1(b) are doubly degenerate.

Next, we lift the degeneracy between the two tunneling terms so that $\tau_{\parallel}$ = 0.165 meV and $\tau_{\perp}$ = 0.145 meV. With these new values we calculate the degree of polarization ($S_{1}$) of the flat-band eigenmodes, obtaining a polarization of 0.128 ($\pm$ depending on the site), in excellent agreement with the experimental value of 0.13. To obtain this value we evaluate, at each k-point, the eigenvectors of $H_{Lieb}^{S}$, which give us the distribution, on the three pillars forming the unit cell, of the H and V populations. This allows the evaluation, for each $k$ point, of the relative H and V population. As a final step, we calculate, for each pillar, the weighted average of the H and V populations on the entire $k$-space and, with these, the polarization on each pillar. (different occupations of s-type flat band modes  can be seen in experimentally measured $E$-$k$ slice [Fig. 2 of the main text]). 
Then, in Fig. S4 we reconstruct the experimental polarization pattern experimentally observed in the S flat-band condensate plotting on each pillar the colour corresponding to its degree of polarization where the emission intensity is above $5\%$, and black (no polarization) where the emission intensity is below this value.
Such a method is justified given in the experiment the polarization degree is measured to be zero in areas where the emission intensity is below signal to noise ratio and hence is purely defined by a background intensity, which is about $5\%$ of the peak signal intensity. Note that our approach assumes an incoherent emission from the different $k$ points. In the condensation regime some degree of phase locking between the amplitudes at different $k$ points may occur, which would result in a weak modulation of intensity distribution across the lattice on a length scale greater than the lattice period. The effect of phase locking is difficult to observe in the experiment due to the condensate covering only 2-3 lattice periods.

\begin{figure}[!htb]
\center
\includegraphics[scale=0.4]{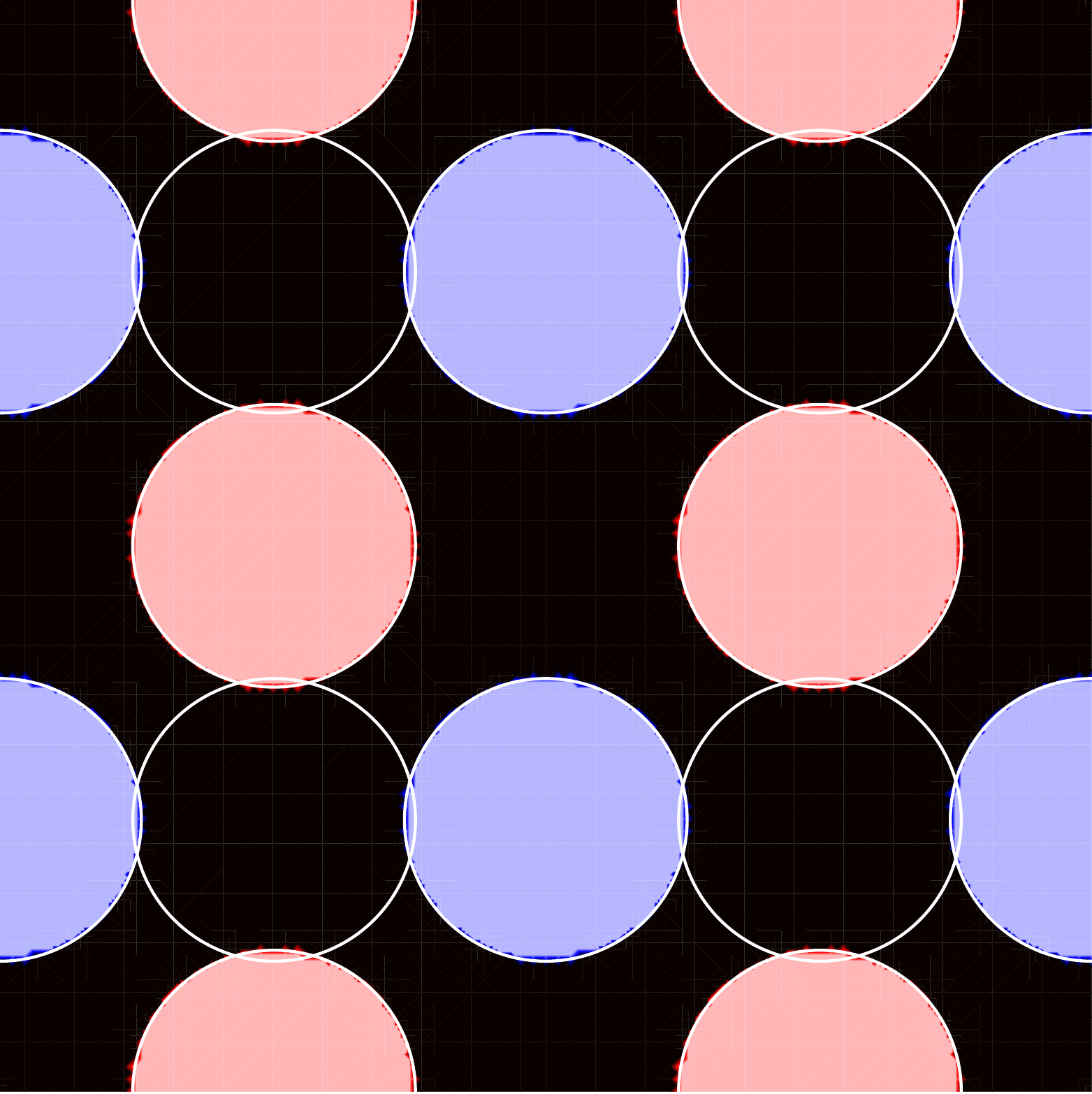}
\includegraphics[scale=0.75]
{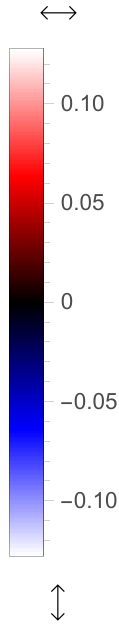}
\center
\vspace{-0.4cm}
\caption{\label{fig-sband}
Polarization pattern for the $S$ band obtained from the TB model. In good agreement with the experiments, $A$ ($C$) sites are horizontally (vertically) polarized, with a degree of polarization of about 0.13.}
\end{figure}

\begin{figure}[!htb]
\center
\includegraphics[scale=0.35]{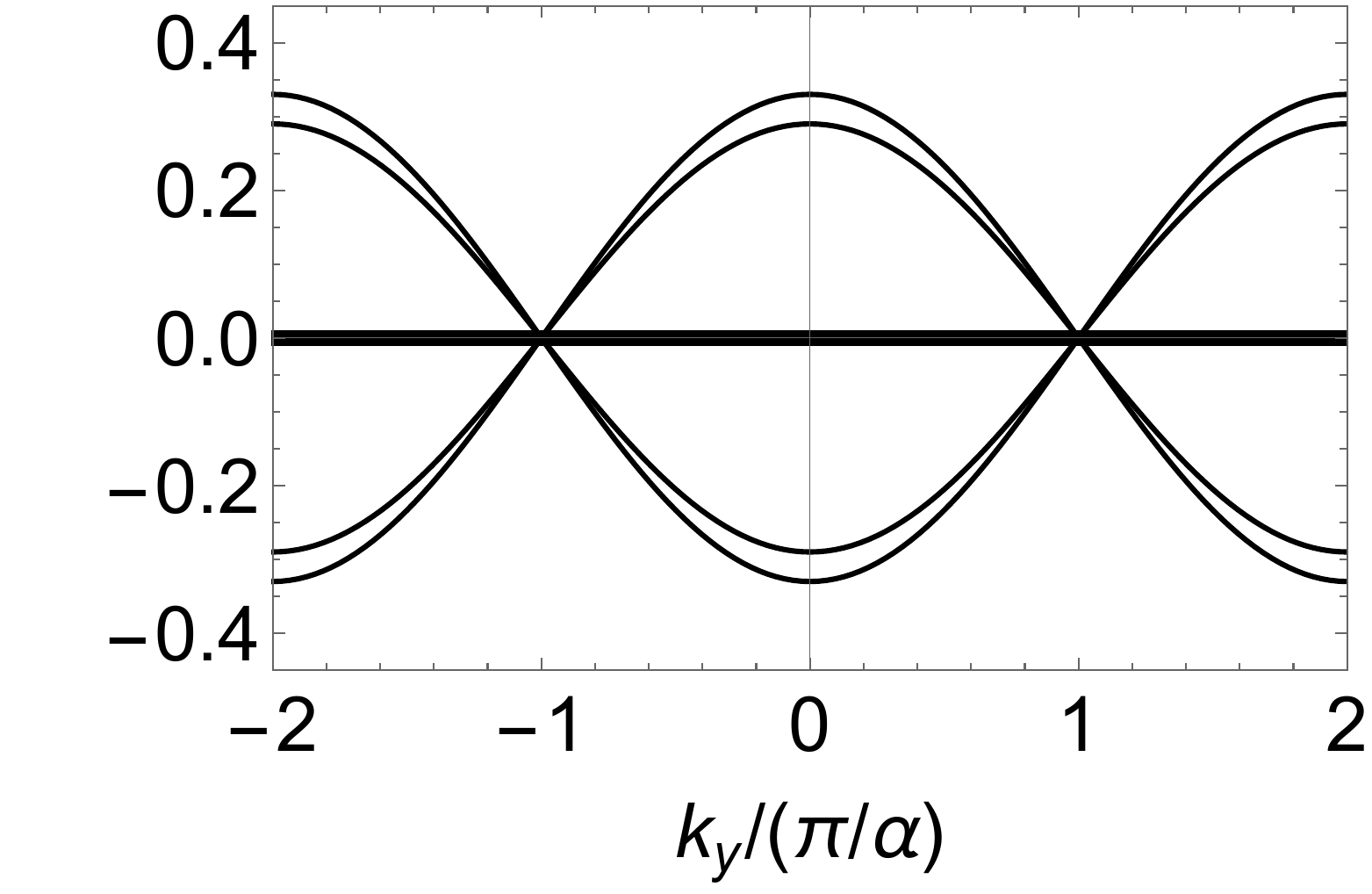}
\includegraphics[scale=0.35]
{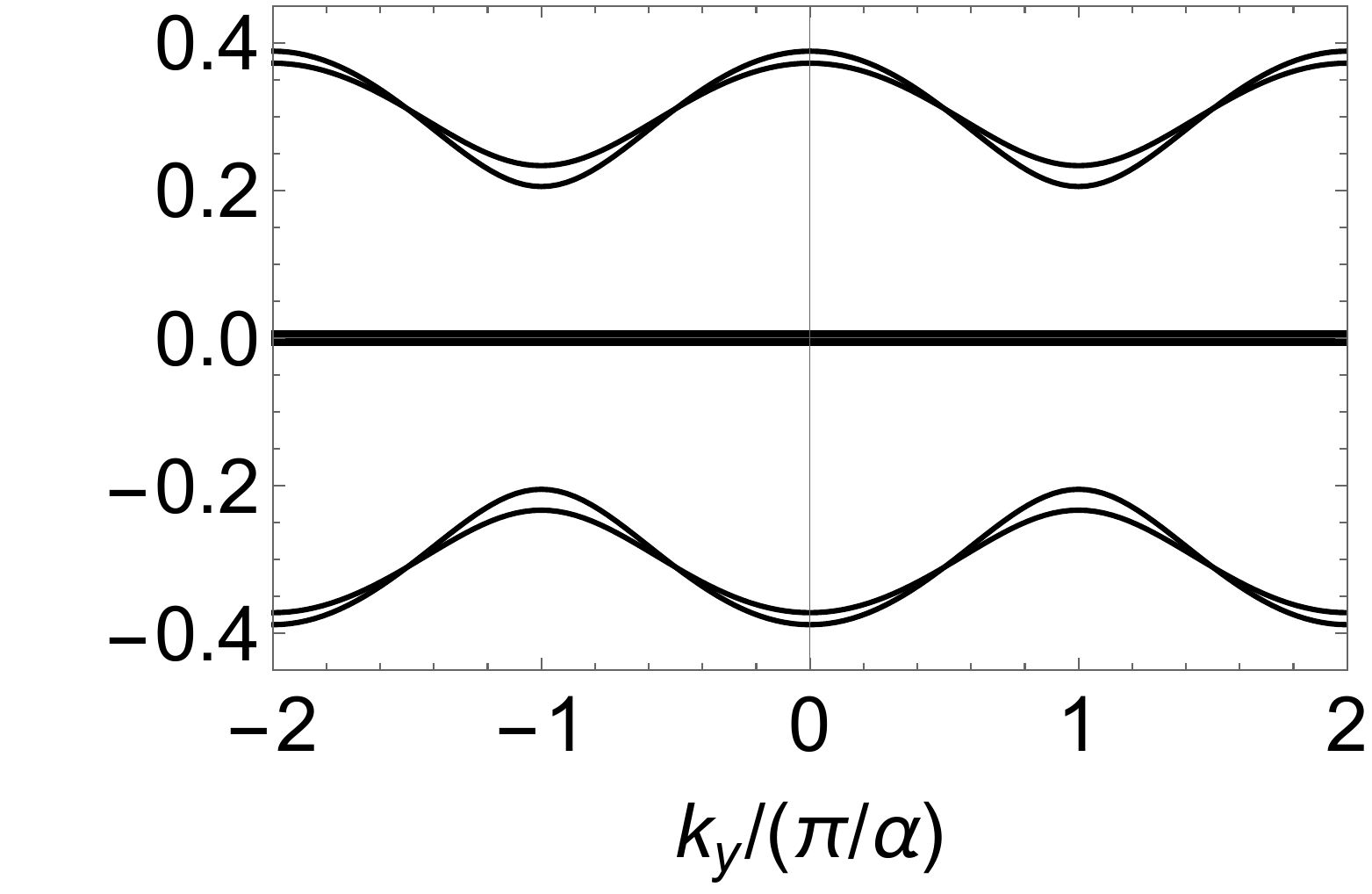}
\includegraphics[scale=0.35]
{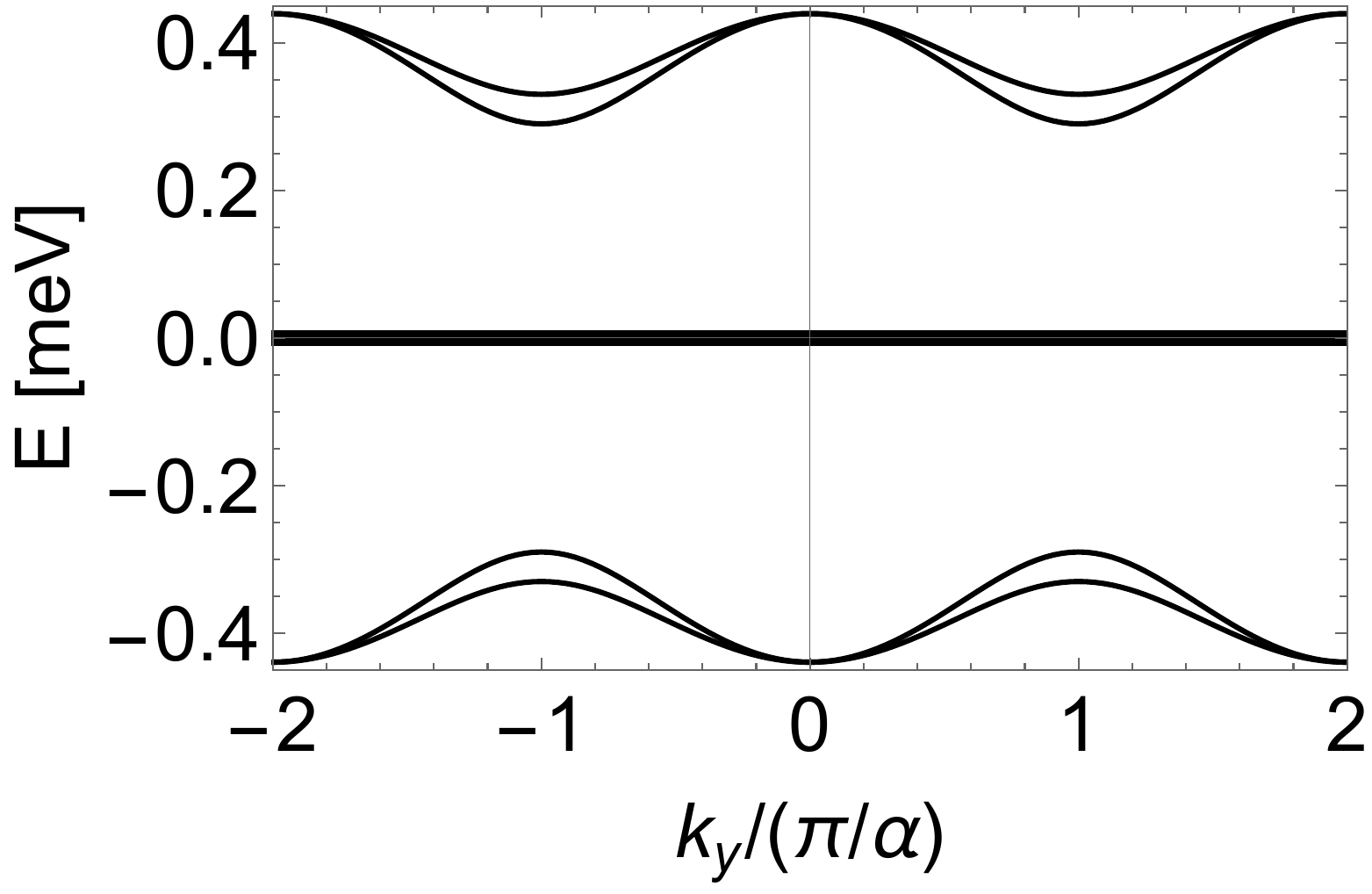}
\caption{\label{fig-band-sband}
\rm{}Calculated $S$ band structure of the Lieb lattice using the TB model for the $S$ modes of each pillar. From left to right $k_x/(\pi/\alpha)= 1, 1.5, 2$. See Fig. \ref{fig-band-pband} for the part of the band obtained from the $P$ modes.}
\end{figure}

\subsection{$P$ band}
Similarly to our treatment of the $S$ bands we define $a_{Hx}$, $a_{Hy}$, $a_{Vx}$, and $a_{Vy}$ as the annihilation operators for the $P$ orbital modes of the pillar with H and V polarization and with the lobes aligned along the $x$ or $y$ directions. The notation follows the same convention for the creation operators and for the modes on the $B$ and $C$ sites. With these definitions, the Hamiltonian for the $P$ band in real space can be written as:

\begin{eqnarray*}
H_{Lieb}^{P}
&=&
-\sum_{m,n=-\infty}^{\infty}b_{Hx_{m,n}}^{\dagger}[\tau_{\perp}^{t}(a_{Hx_{m,n}}+a_{Hx_{m-1,n}})-\tau_{\parallel}^{a}(c_{Hx_{m,n}}+c_{Hx_{m,n-1}})]
\\
&&
-\sum_{m,n=-\infty}^{\infty}b_{Hy_{m,n}}^{\dagger}[\tau_{\perp}^{a}(a_{Hy_{m,n}}+a_{Hy_{m-1,n}})-\tau_{\parallel}^{t}(c_{Hy_{m,n}}+c_{Hy_{m,n-1}})]
\\
&&
-\sum_{m,n=-\infty}^{\infty}b_{Vx_{m,n}}^{\dagger}[\tau_{\parallel}^{t}(a_{Vx_{m,n}}+a_{Vx_{m-1,n}})-\tau_{\perp}^{a}(c_{Vx_{m,n}}+c_{Vx_{m,n-1}})]
\\
&&
-\sum_{m,n=-\infty}^{\infty}b_{Vy_{m,n}}^{\dagger}[\tau_{\parallel}^{a}(a_{Vy_{m,n}}+a_{Vy_{m-1,n}})-\tau_{\perp}^{t}(c_{Vy_{m,n}}+c_{Vy_{m,n-1}})]\,+h.c.,
\end{eqnarray*}

\noindent
where the hopping probabilities $\tau_{\parallel}^{a}$, $\tau_{\parallel}^{t}$, $\tau_{\perp}^{a}$, and $\tau_{\perp}^{t}$ correspond to modes hopping from one site to another having the polarization parallel $(\parallel)$ or perpendicular $(\perp)$ to the hopping direction and the lobes of the $P$ orbital aligned $(a)$ or transverse $(t)$ to the hopping direction. As before one can introduce the Fourier transform of the creation and annihilation operators and diagonalize the Hamiltonian in $k$-space. This time the kernel matrix will be $12\times12$ since there are 2 modes with 2 possible polarizations on each of the 3 pillars. In this case the Hamiltonian can be written in a compact form as:

\begin{equation*}
H_{Lieb}^{P}=-2\sum_{p,q=-\infty}^{\infty}\psi_{p,q}^{P\dagger^T}
\begin{pmatrix}
M_{Hx,p,q}^P & 0 & 0 & 0 \\ 0 & M_{Hy,p,q}^P & 0 & 0 \\ 0 & 0 & M_{Vx,p,q}^P & 0 \\ 0 & 0 & 0 & M_{Vy,p,q}^P 
\end{pmatrix}\psi_{p,q}^P\,\,,
\end{equation*}

\noindent
with

\begin{equation*}
\psi_{p,q}^{P\dagger^T}=(
a^{\dagger}_{Hx,p,q} ,
a^{\dagger}_{Hy,p,q} ,
b^{\dagger}_{Hx,p,q} ,
b^{\dagger}_{Hy,p,q} ,
c^{\dagger}_{Hx,p,q} , c^{\dagger}_{Hy,p,q} , a^{\dagger}_{Vx,p,q} ,
a^{\dagger}_{Vy,p,q} ,
b^{\dagger}_{Vx,p,q} ,
b^{\dagger}_{Vy,p,q} ,
c^{\dagger}_{Vx,p,q} , c^{\dagger}_{Vy,p,q}
),
\end{equation*}

\noindent
and

\begin{equation*}
M_{Hx,p,q}^P=
\begin{pmatrix}
0 & \tau_{\perp}^te^{+\frac{i\alpha k_q}{2}}\cos\left(\frac{k_q\alpha}{2}\right) & 0 \\
\tau_{\perp}^{t\star}e^{-\frac{i\alpha k_q}{2}}\cos\left(\frac{k_q\alpha}{2}\right)  & 0 & \tau_{\parallel}^ae^{+\frac{i\alpha k_p}{2}}\cos\left(\frac{k_p\alpha}{2}\right) \\
0 & \tau_{\parallel}^{a\star}e^{-\frac{i\alpha k_p}{2}}\cos\left(\frac{k_p\alpha}{2}\right) & 0 
\end{pmatrix},
\end{equation*}

\begin{equation*}
M_{Hy,p,q}^P=
\begin{pmatrix}
0 & \tau_{\perp}^ae^{+\frac{i\alpha k_q}{2}}\cos\left(\frac{k_q\alpha}{2}\right) & 0 \\
\tau_{\perp}^{a\star}e^{-\frac{i\alpha k_q}{2}}\cos\left(\frac{k_q\alpha}{2}\right)  & 0 & \tau_{\parallel}^te^{+\frac{i\alpha k_p}{2}}\cos\left(\frac{k_p\alpha}{2}\right) \\
0 & \tau_{\parallel}^{t\star}e^{-\frac{i\alpha k_p}{2}}\cos\left(\frac{k_p\alpha}{2}\right) & 0 
\end{pmatrix},
\end{equation*}

\begin{equation*}
M_{Vx,p,q}^P=
\begin{pmatrix}
0 & \tau_{\parallel}^te^{+\frac{i\alpha k_q}{2}}\cos\left(\frac{k_q\alpha}{2}\right) & 0 \\
\tau_{\parallel}^{t\star}e^{-\frac{i\alpha k_q}{2}}\cos\left(\frac{k_q\alpha}{2}\right)  & 0 & \tau_{\perp}^ae^{+\frac{i\alpha k_p}{2}}\cos\left(\frac{k_p\alpha}{2}\right) \\
0 & \tau_{\perp}^{a\star}e^{-\frac{i\alpha k_p}{2}}\cos\left(\frac{k_p\alpha}{2}\right) & 0 
\end{pmatrix},
\end{equation*}

\begin{equation*}
M_{Vx,p,q}^P=
\begin{pmatrix}
0 & \tau_{\parallel}^ae^{+\frac{i\alpha k_q}{2}}\cos\left(\frac{k_q\alpha}{2}\right) & 0 \\
\tau_{\parallel}^{a\star}e^{-\frac{i\alpha k_q}{2}}\cos\left(\frac{k_q\alpha}{2}\right)  & 0 & \tau_{\perp}^te^{+\frac{i\alpha k_p}{2}}\cos\left(\frac{k_p\alpha}{2}\right) \\
0 & \tau_{\perp}^{t\star}e^{-\frac{i\alpha k_p}{2}}\cos\left(\frac{k_p\alpha}{2}\right) & 0 
\end{pmatrix}.
\end{equation*}

As before, the polarized eigenvector of the $P$ band can be easily written in terms of the function $E[x,y]=-2xe^{-\frac{i\alpha y}{2}}cos\left(\frac{y\alpha}{2}\right)$, where $x$ is the hopping probability and $y$ the direction of propagation:

\begin{eqnarray*}
e_1&=&\frac{1}{A_1}\left\{0,-\frac{E[\tau^t_{\perp},k_q]}{E[\tau^a_{\parallel},k_p]},1,0,0,0,0,0,0,0,0,0\right\}\\
e_2&=&\frac{1}{A_2}\left\{-\frac{\sqrt{|E[\tau^a_{\parallel},k_p]|^2+|E[\tau^t_{\perp},k_q]|^2}}{E[\tau^t_{\perp},k_q]^*},\frac{E[\tau^a_{\parallel},k_p]^*}{E[\tau^t_{\perp},k_q]^*},1,0,0,0,0,0,0,0,0,0\right\}\\
e_3&=&\frac{1}{A_3}\left\{+\frac{\sqrt{|E[\tau^a_{\parallel},k_p]|^2+|E[\tau^t_{\perp},k_q]|^2}}{E[\tau^t_{\perp},k_q]^*},\frac{E[\tau^a_{\parallel},k_p]^*}{E[\tau^t_{\perp},k_q]^*},1,0,0,0,0,0,0,0,0,0\right\}\\
e_4&=&\frac{1}{A_4}\left\{0,0,0,0,-\frac{E[\tau^a_{\perp},k_q]}{E[\tau^t_{\parallel},k_p]},1,0,0,0,0,0,0\right\}\\
e_5&=&\frac{1}{A_5}\left\{0,0,0,-\frac{\sqrt{|E[\tau^t_{\parallel},k_p]|^2+|E[\tau^a_{\perp},k_q]|^2}}{E[\tau^a_{\perp},k_q]^*},\frac{E[\tau^t_{\parallel},k_p]^*}{E[\tau^a_{\perp},k_q]^*},1,0,0,0,0,0,0\right\}\\
e_6&=&\frac{1}{A_6}\left\{0,0,0,+\frac{\sqrt{|E[\tau^t_{\parallel},k_p]|^2+|E[\tau^a_{\perp},k_q]|^2}}{E[\tau^a_{\perp},k_q]^*},\frac{E[\tau^t_{\parallel},k_p]^*}{E[\tau^a_{\perp},k_q]^*},1,0,0,0,0,0,0\right\}\\
e_7&=&\frac{1}{A_7}\left\{0,0,0,0,0,0,0,-\frac{E[\tau^t_{\parallel},k_q]}{E[\tau^a_{\perp},k_p]},1,0,0,0\right\}\\
e_8&=&\frac{1}{A_8}\left\{0,0,0,0,0,0,-\frac{\sqrt{|E[\tau^t_{\parallel},k_q]|^2+|E[\tau^a_{\perp},k_p]|^2}}{E[\tau^t_{\parallel},k_q]^*},\frac{E[\tau^a_{\perp},k_p]^*}{E[\tau^t_{\parallel},k_q]^*},1,0,0,0\right\}\\
e_9&=&\frac{1}{A_9}\left\{0,0,0,0,0,0,+\frac{\sqrt{|E[\tau^t_{\parallel},k_q]|^2+|E[\tau^a_{\perp},k_p]|^2}}{E[\tau^t_{\parallel},k_q]^*},\frac{E[\tau^a_{\perp},k_p]^*}{E[\tau^t_{\parallel},k_q]^*},1,0,0,0\right\}\\
e_{10}&=&\frac{1}{A_{10}}\left\{0,0,0,0,0,0,0,0,0,0,-\frac{E[\tau^a_{\parallel},k_q]}{E[\tau^t_{\perp},k_p]},1\right\}\\
e_{11}&=&\frac{1}{A_{11}}\left\{0,0,0,0,0,0,0,0,0,-\frac{\sqrt{|E[\tau^a_{\parallel},k_q]|^2+|E[\tau^t_{\perp},k_p]|^2}}{E[\tau^a_{\parallel},k_q]^*},\frac{E[\tau^t_{\perp},k_p]^*}{E[\tau^a_{\parallel},k_q]^*},1\right\}\\
e_{12}&=&\frac{1}{A_{12}}\left\{0,0,0,0,0,0,0,0,0,+\frac{\sqrt{|E[\tau^a_{\parallel},k_q]|^2+|E[\tau^t_{\perp},k_p]|^2}}{E[\tau^a_{\parallel},k_q]^*},\frac{E[\tau^t_{\perp},k_p]^*}{E[\tau^a_{\parallel},k_q]^*},1\right\}
\end{eqnarray*}

\noindent where the twelve constants $A_i$ are normalization constants. As for the $S$ band, we first neglect the polarization degree of freedom (i.e. $\tau_{\parallel}^a=\tau_{\perp}^a=\tau^a$ and $\tau_{\parallel}^t=\tau_{\perp}^t=\tau^t$) and fit the experimental $P$ band [see Fig. 1(b) in the main text and Fig. \ref{fig-both-bands}] to obtain the values $\tau^a=0.375$ meV and $\tau^t=0.100$ meV. Next we introduce the polarization dependence: $\tau_{\parallel}^a=0.375$, $\tau_{\perp}^a=0.125$, $\tau_{\parallel}^t=0.100$ and $\tau_{\perp}^t=0.033$ meV. With these values we calculate the degree of polarization ($S_1$) and the ratio between $P_x$ and $P_y$ orbitals of the $P$ flat-band eigenmodes. We obtain a polarization degree of the order of 0.4 ($\pm$ depending on the sites) and ratio between $P_x$ and $P_y$ orbitals on A sites of 3.7 (the inverse applies to $C$ sites), in good agreement with the experimental results. To obtain these values we evaluate, at each k-point, the eigenvectors of $H_{Lieb}^{P}$, which give us the distribution, on the three pillars forming the unit cell, of the H and V populations. This allows the evaluation, for each k-point, of the relative H and V population, and therefore the degree of polarization, on each pillar. Note that, with respect to the S-band case, these eigenvalues also give us the relative intensity of the $P_x$ and $P_y$ orbitals. As a final step, we evaluate, for each pillar, the average polarization, and the average relative $P_x$ and $P_y$ populations on the entire k-space. As for the $S$ band, we calculate, for each pillar, the weighted average of the H and V populations and of the $P_x$ and $P_y$ populations on the entire $k$-space and, with these, the polarization on each pillar and the ratio between the two $P$-like orbitals (different occupations of modes in the $P$-type flat band above threshold can be seen in experimentally measured $E$-$k$ slice [Fig. 2 of the main text]). 
Then, in Fig. S6 we reconstruct the experimental population distribution and polarization pattern experimentally observed in the P flat-band condensate. For the population distribution, we use two Hermite-Gauss $P_x$ and $P_y$ modes with the relative populations just found, for the polarization we plot on each pillar the colour corresponding to its degree of polarization where the emission intensity is above $5\%$, and black (no polarization) where the emission intensity is below this value similarly to the case of s-type flat band condensates.

\begin{figure}[!h]
\center
\includegraphics[scale=0.35]{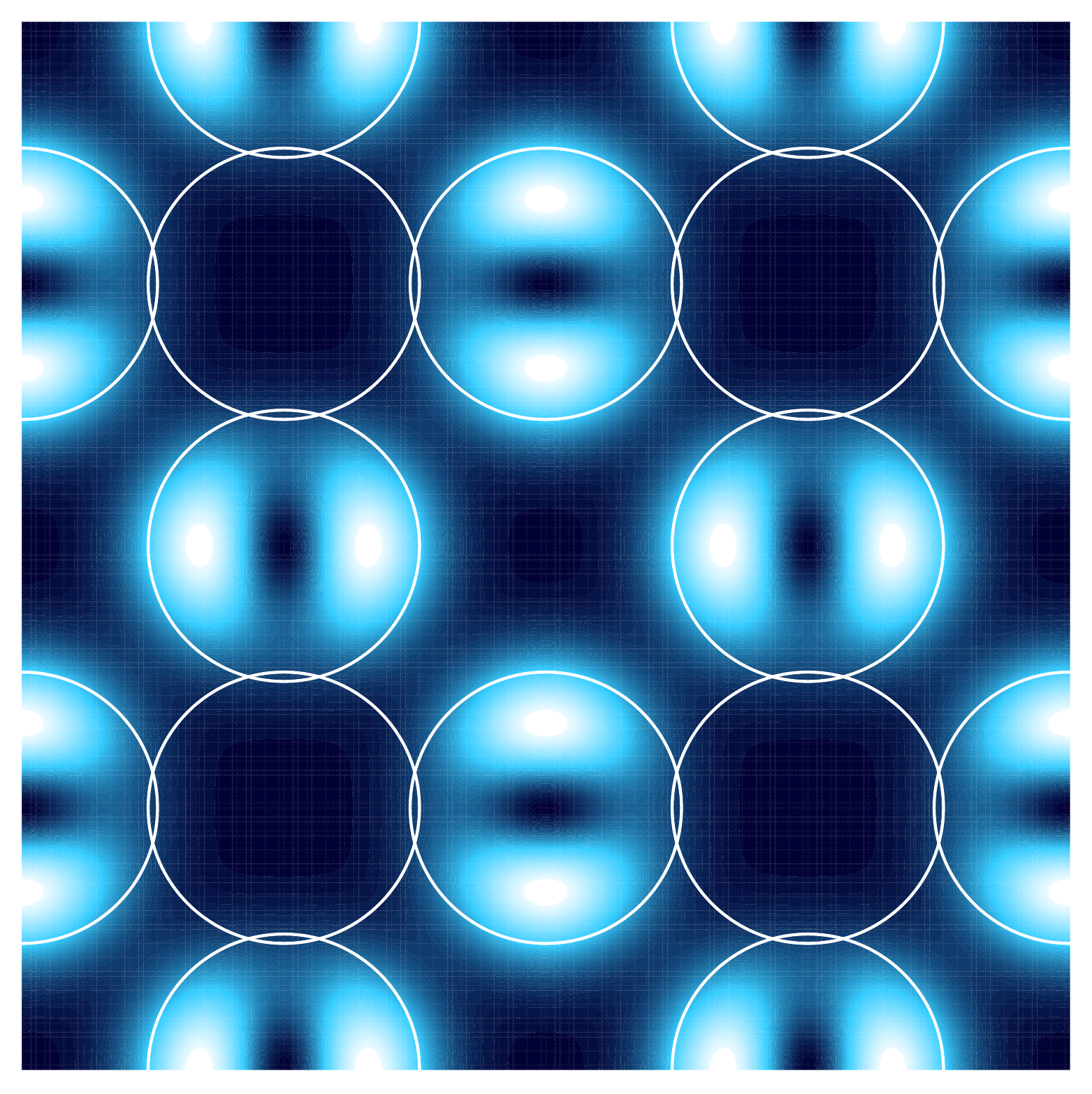}
\includegraphics[scale=0.35]
{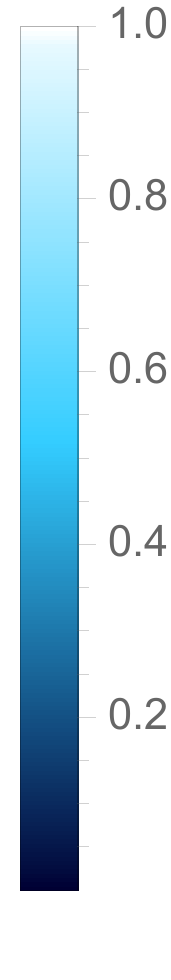}
\includegraphics[scale=0.35]{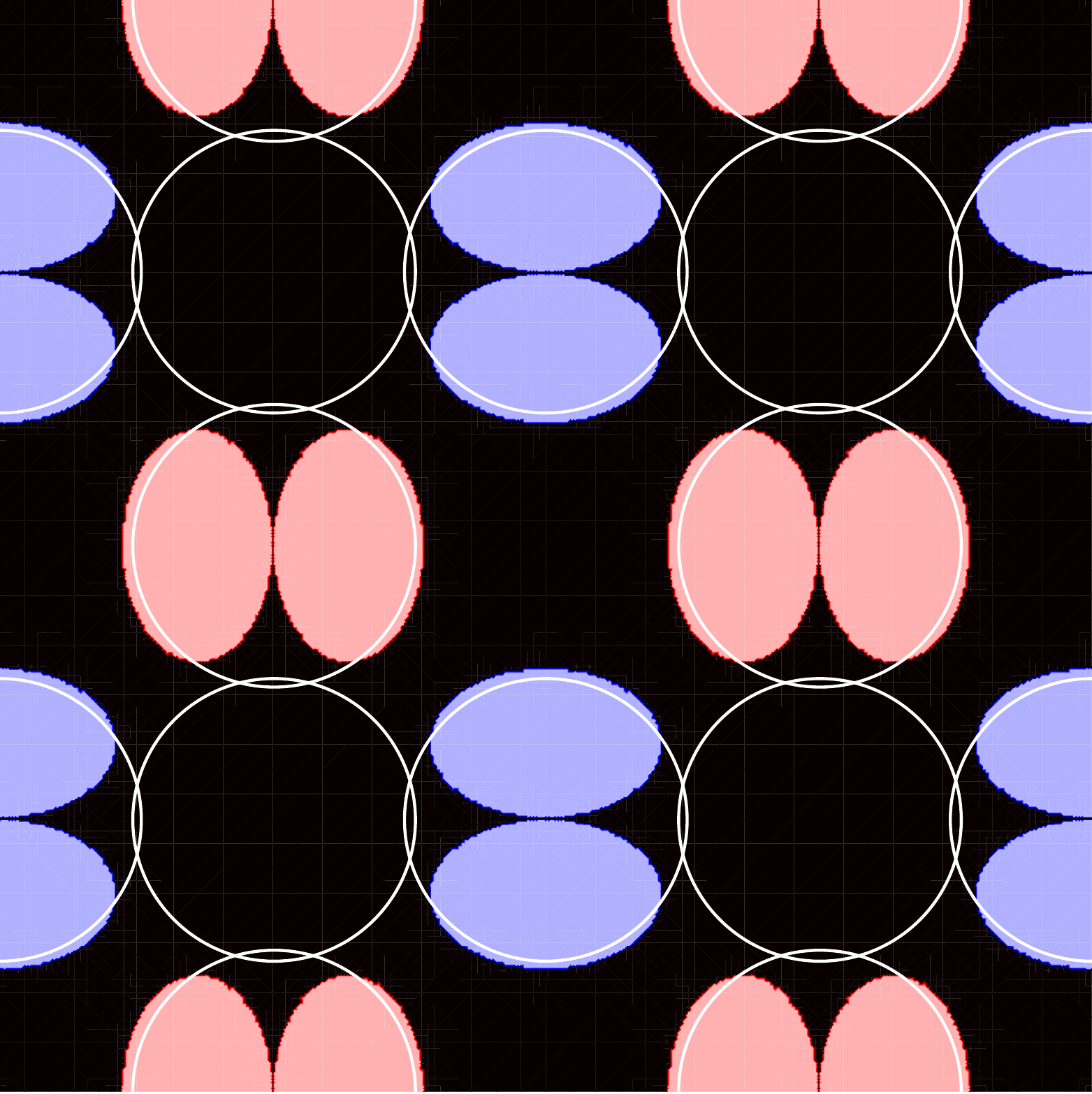}
\includegraphics[scale=0.6]
{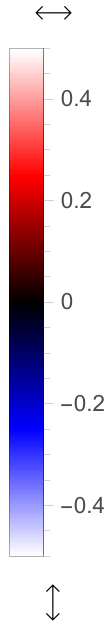}
\center
\vspace{-0.4cm}
\caption{\label{fig-pband}
Calculated real space emission (left) and polarization pattern (right) for the $P$ flat band obtained from the TB model. In good agreement with the experiments, on the A (C) sites the polariton population is predominantly (with a ratio of 5.9) in $P_x$ ($P_y$) orbitals and is horizontally (vertically) polarized with a polarization degree of 0.43.}
\end{figure}

\begin{figure}[!ht]
\center
\includegraphics[scale=0.35]{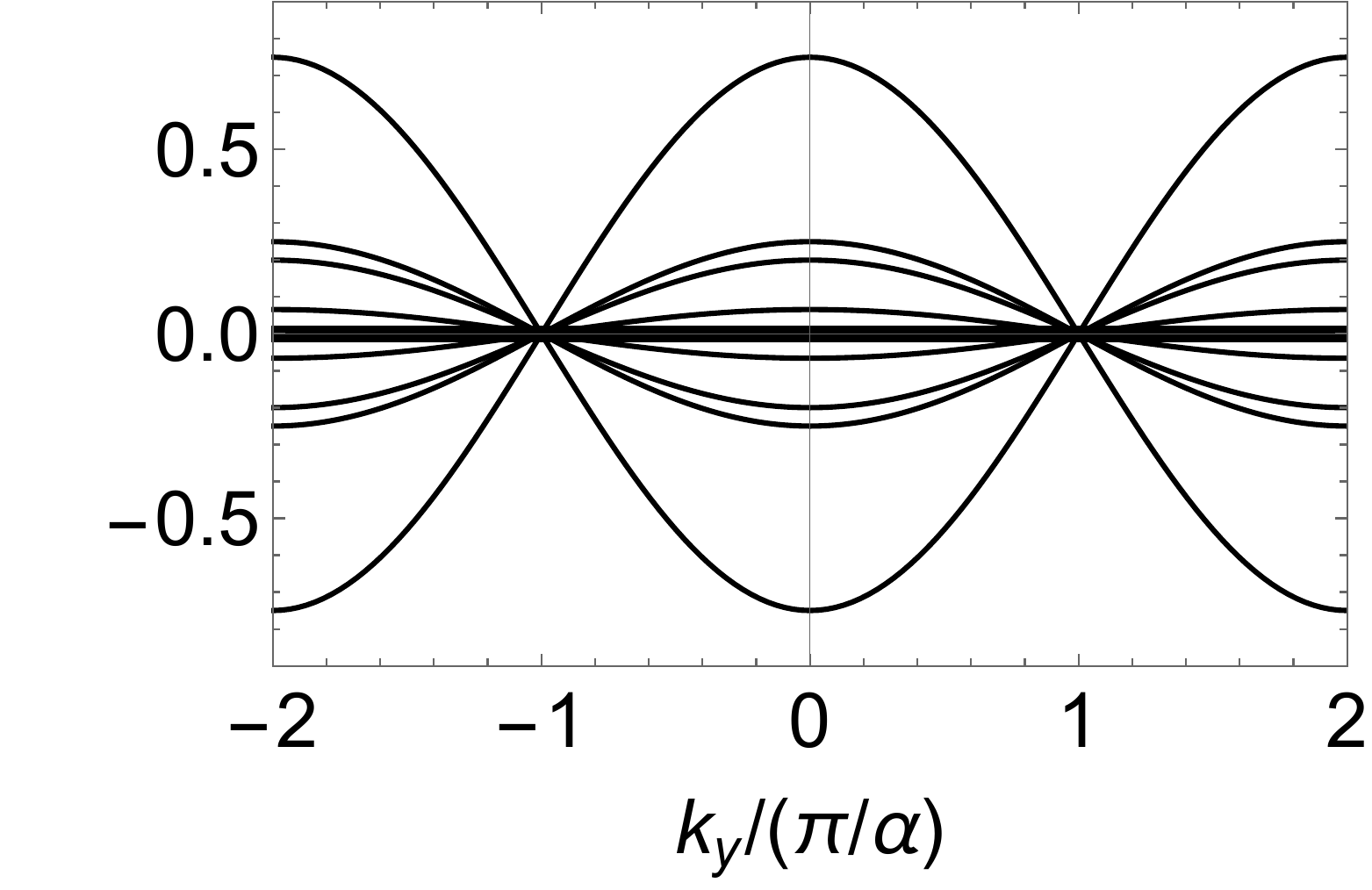}
\includegraphics[scale=0.35]
{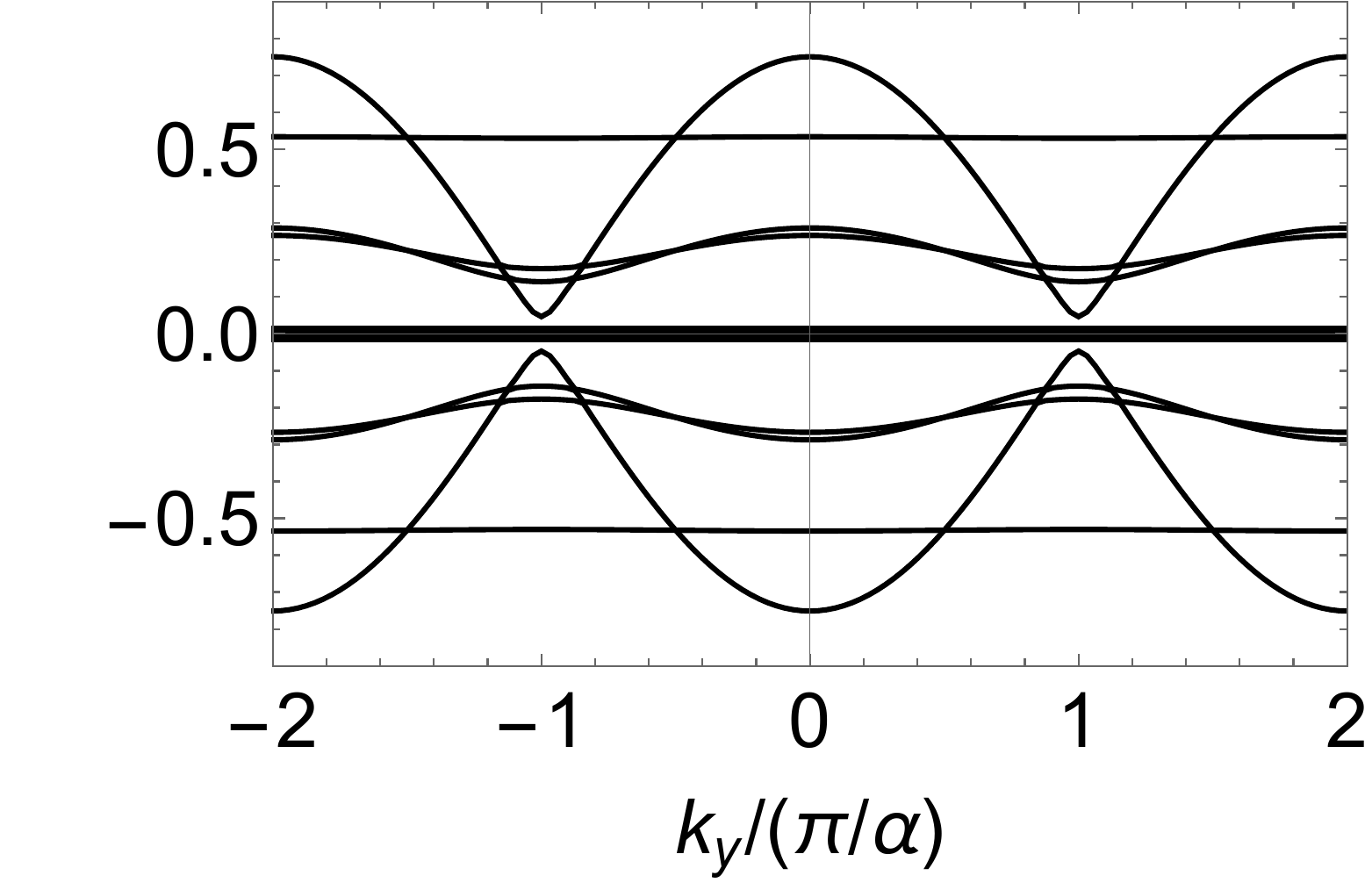}
\includegraphics[scale=0.35]
{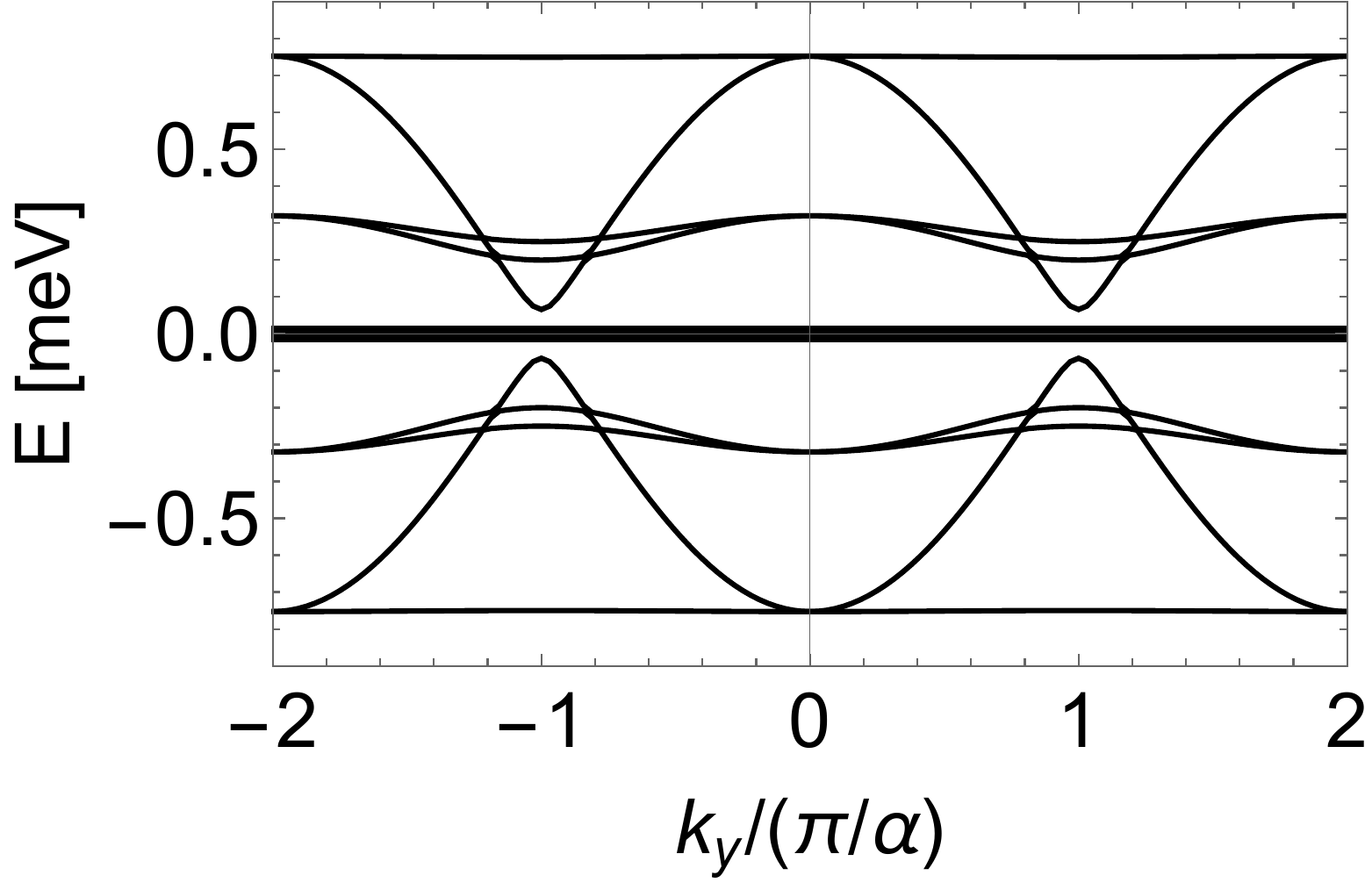}
\caption{\label{fig-band-pband}
\rm{}Calculated band structure of the Lieb lattice using the TB model for the $P$ modes of each pillar. From left to right $k_x/(\pi/\alpha)= 1, 1.5, 2$. See Fig. \ref{fig-band-sband} for the part of the band obtained from the $S$ modes.}
\end{figure}

\section{2D Schr\"{o}dinger equation in a periodic potential}\label{cm}

\begin{figure}[!htb]
\center
\includegraphics[width=5.0in]{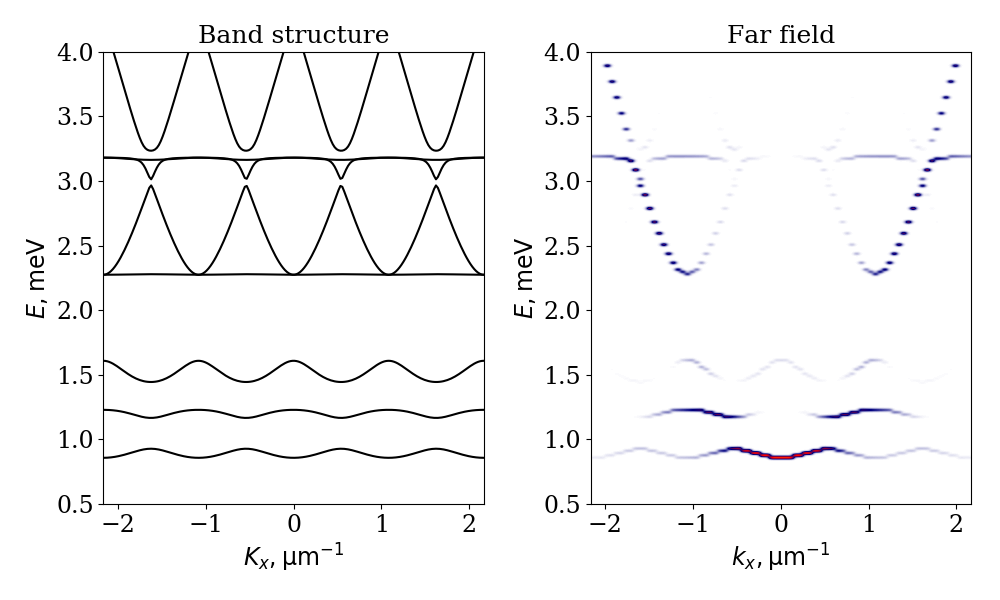}
\center
\caption{\label{cell} Band structure for Lieb lattice at $k_{y} = 0$ and the expected far-field emission.}
\end{figure}

In addition to the TB model described in the previous section, we performed a detailed numerical study using the Schr\"{o}dinger equation in a Lieb lattice potential in order to explain the missing bands of the experimental band structure [see Fig. 1(b) of the main text]. Moreover, in order to better reproduce the experimental data, we also include the effect of a finite lifetime. In contrast to the TB model described above this model allows the reproduction of the band structure and polariton emission in energy-momentum space essentially without fitting parameters. First, we find the Bloch states and band structure of the Lieb lattice. For this we solve the stationary Schr\"{o}dinger equation on a discretized Lieb lattice cell with potential $U(\mathbf{r})$ formed by microcavity pillars in a Lieb lattice. Inside the bulk of our lattice the top DBR was only partially etched without affecting the active region [Fig. \ref{lattice}(d)]. Whilst the depth of the etching is not exactly known we estimate that about 4-6 DBR pairs in the top mirror are left after etching [Fig. \ref{lattice}(d)]. Using the transfer matrix method we estimate the potential landscape created by our etched pillars, which is in the form of a circular well:
$$
U(\mathbf{r})=
\begin{cases}
0,~\text{ inside pillar}\\
10\;{\rm meV},~\text{ outside pillar}
\end{cases}
$$
We take effective mass $m^*=5\times 10^{-5} m_e$ estimated from the experimental data for polaritons in an unconstrained cavity [shown in Fig. \ref{lattice}(a)].
The calculated band structure is shown in Fig.~\ref{cell}.

We also calculate the far-field emission as a function of the wavevector $\mathbf{k}$ of the outgoing radiation. We assume all Bloch states in the Brillouin zone (BZ) are equally populated. Then, the histogram for the far-field emission can be calculated as a sum of intensities contributed by each Bloch wave function $\Psi_{\mathbf{K},n}(\mathbf{r})$ of the first BZ at a given energy, that is, 
$$
I(\mathbf{k},E)\sim \sum_{E_n(\mathbf{K})=E} |F[\Psi_{\mathbf{K},n}(\mathbf{r})]|^2
$$
where $F$ denotes the Fourier transform and the quasimomentum $\mathbf{K}$ belongs to the first BZ. The predicted far-field intensity is presented in Fig.~\ref{cell} and is similar to that observed in our experiments.

\begin{figure*}[h!]
\begin{center}
\includegraphics[scale = 0.6]{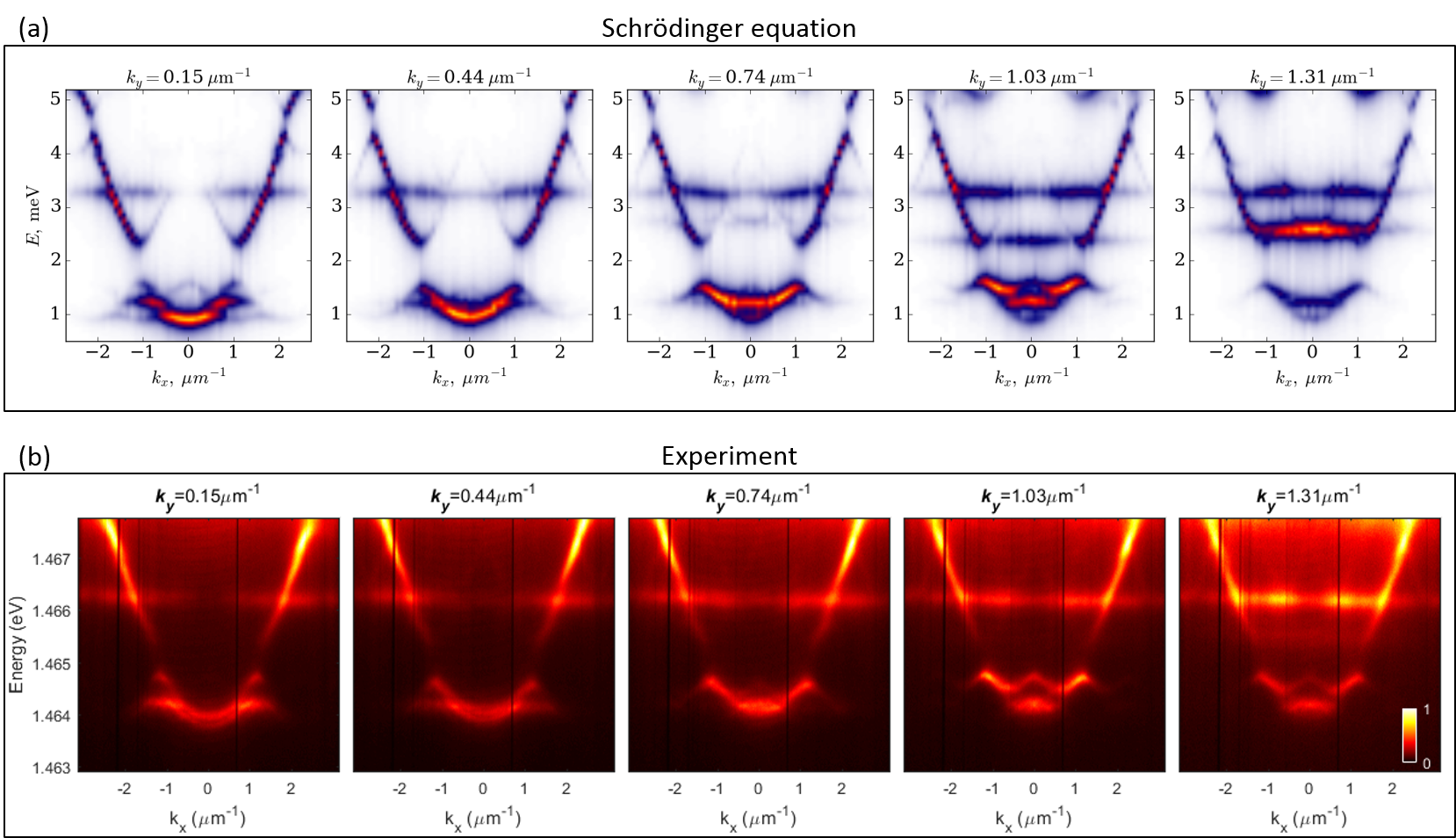}
\caption{\label{damped}
Experimentally measured (a) and numerically calculated (b) far-field emission of the Lieb lattice at $k_{y}/(\pi/\alpha)$ = 0.3, 0.8, 1.4, 1.9, 2.4.
}
\end{center}
\end{figure*}

The far-field emission presented in Fig.~\ref{cell} does not take into account the finite lifetime of polaritons in the real system. To account for the damping effects we solve the time-dependent Schr\"{o}dinger equation
\begin{equation}
i\hbar\frac{\partial\Psi}{\partial t} = -\frac{\hbar^2}{2m^*}\Delta \Psi + U(\mathbf{r})\,\Psi
\label{Sch}
\end{equation}
where the real part of the potential ${\rm Re}\; U(\mathbf{r})$
is a Lieb lattice formed by microcavity pillars and ${\rm Im}\; U(\mathbf{r})$ accounts for the finite polariton lifetime. We take 
${\rm Re}\; U(\mathbf{r})$ as in the band structure calculation (a 10 meV well) and ${\rm Im}\; U(\mathbf{r})$ equal to $-0.1$ inside pillars and $-0.5$ outside pillars. We take random initial conditions in the form
\begin{equation}
\Psi(\mathbf{r},0)=\sum_{\mathbf{K}\in BZ} \sum_{n} c_{\mathbf{K},n} \Psi_{\mathbf{K},n}(\mathbf{r})
\label{ic}
\end{equation}
where random complex coefficients $c_{\mathbf{K},n}$ satisfying $|c_{\mathbf{K},n}|=1$ and $\Psi_{\mathbf{K},n}(\mathbf{r})$ are Bloch wave functions found from the stationary analysis described above.
Starting from random initial conditions~\eqref{ic} we evolve the time-dependent Schr\"{o}dinger equation~\eqref{Sch} for time sufficient for polaritons to decay due to the inherent damping.
Initial conditions~\eqref{ic} imply that all zones are excited homogeneously for each $\mathbf{K}$ in the first BZ, as in the previous calculation.
The resulting intensity in the far-field is given by the average
over the random initial phases 
$I(\mathbf{k},E)\sim \langle |F[\Psi(\mathbf{r},t)]|^2 \rangle_{c_{\mathbf{k},n}}$
where $F$ denotes the Fourier transform in time and spatial coordinates. Examples of the far-field emission in the presence of damping for different fixed in-plane wave vectors is shown in Fig.~\ref{damped}(b). The results are consistent with the experimentally measured spectra shown in Fig.~\ref{damped}(a), where we see highly inhomogeneous emission intensity across the full set of energy bands calculated and shown in Fig. \ref{cell}.

\section{Estimation of $P_{x,y}$ orbital populations}

\begin{figure}[h!]
\center
\includegraphics[scale=0.4]{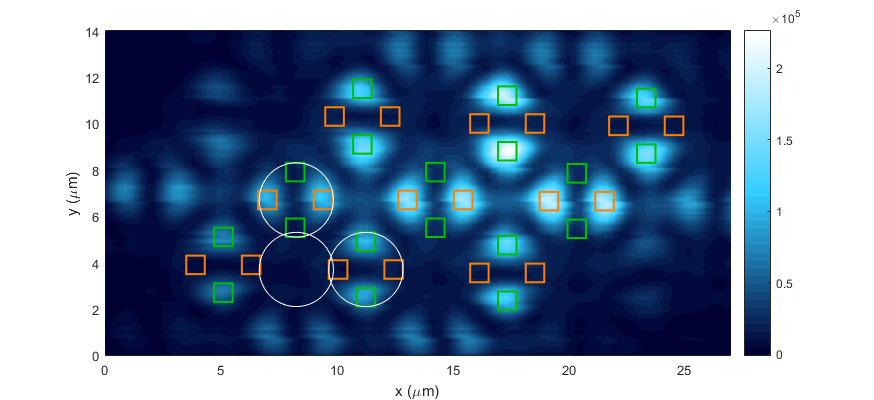}
\center
\vspace{-0.4cm}
\caption{The spatial distribution of the transmitted intensity at the energy of the $P$ flat band (corresponding to Fig. 2(j) of the main text). The orange and green squares indicate the integration area used to evaluate the polariton population attributed to $P_{x}$ and $P_{y}$ orbitals, respectively, in the main text. The white circles show pillars constituting one unit cell.}
\label{databinning}
\end{figure}

Fig. \ref{databinning} shows the data binning technique used to estimate the relative populations of the two orthogonal $P$ orbitals in the $P$ flat-band condensate. The intensity (CCD counts) detected within the left/right and top/bottom squares delimited by the orange and green squares are attributed to the $P_{x}$ and $P_{y}$ orbitals respectively. For each pillar the ratio of the populations of the more intense orbital to that of the less intense orbital was taken, corresponding to $|\psi_{P_{x}}|^2/|\psi_{P_{y}}|^2$ and $|\psi_{P_{y}}|^2/|\psi_{P_{x}}|^2$ for the $A$ and $C$ sites respectively. The mean of the ratios gives a factor of 6.2 for the populations between the two orbitals.  

\section{Energy fragmentation of flat band condensates}

Above the threshold for multimode condensation in our sample, the microscopic spectral properties of the polariton condensates were studied using the protocol developed in Ref. \cite{PhysRevLett.116.066402} as shown in the main text. This required analysis of the energy-resolved slices (corresponding to a narrow line of real space on the sample) of the tomographic images taken by scanning the final lens before the spectrometer. By looking along lines of pillars, peaks in the real-space energy spectrum were identified and attributed to underlying photonic orbitals. 

\begin{figure}[h!]
\center
\includegraphics[scale=1.2]{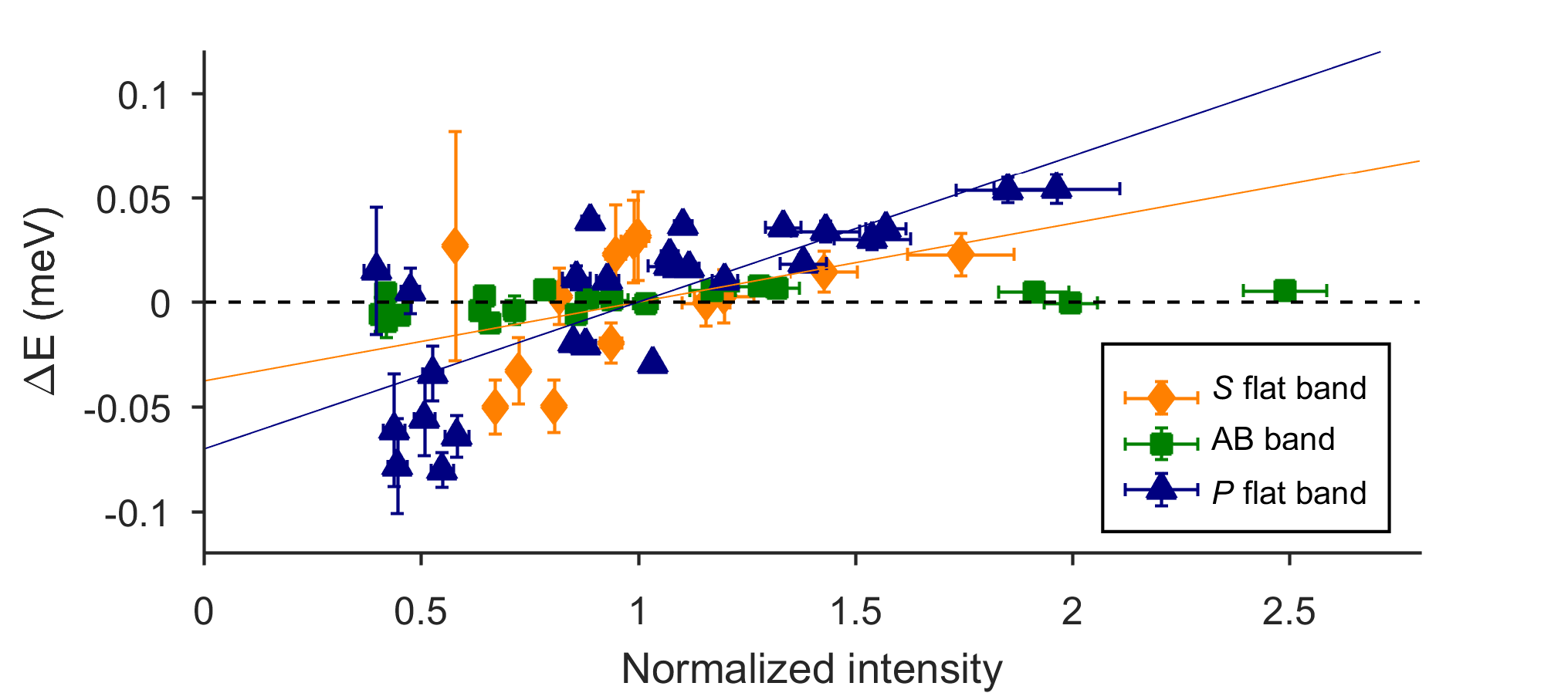}
\center
\caption{\label{fig1} On-site energy (with respect to ensemble average) vs. normalized intensity of individual photonic orbitals for data shown in Fig. 3 of main text.}
\label{frag}
\end{figure}

In Fig. \ref{frag} we show a plot of the data presented in Fig. 3 of the main text. Each data point corresponds to the emission from one orbital lobe (1 and 2 bright lobes for the $S$ and $P$ orbitals respectively). Along the horizontal axis we see the intensity of the identified peaks, normalized to the average intensity of the ensemble (the Gaussian pump creates a broad distribution of intensities across the sites). Along the vertical axis the energy detuning of the peaks relative to the average condensate energy is plotted. A significant energy variation about the mean can be seen in the case of the flat-band condensates, which is absent for the AB band condensates. A positive correlation (positive slope) between the population 
(intensity) and on-site energy exists for the flat bands, as demonstrated by the solid fitted least squares lines, whereas the AB band data points all lie at the mean energy (dashed black line). Note that the energy fragmentation is on the order of 100-150 $\mu$eV, which lies within the spectral width of the flat-band condensates ($\sim 0.2$ meV) hence enabling the destructive interference associated with flat bands.  

In Ref. \cite{PhysRevLett.116.066402}, Baboux and co-workers reported a similar fragmentation effect, which they attribute to photonic disorder in the sample and that arising from imperfections of the fabrication process. Here we estimate that the disorder is up to a few tens of $\mu$eV, whilst the fragmentation shown in Fig. \ref{frag} exceeds 100 $\mu$eV. Furthermore we demonstrate a clear correlation between the energy of the condensate emission on individual lattice sites and the population \footnote{We use a standard bootstrap method and calculate the Pearson correlation coefficient $r$ between the normalized intensity and $\Delta$E for both the $S$ and $P$ flat bands, taking 1000 resamples of the data. This yields values of $r = 0.4139$ and $r = 0.7672$ for the $S$ and $P$ flat bands respectively.}, suggesting that the effect of interactions on the condensate fragmentation is dominant. 

\section{Discussion of strong vs. weak coupling}

Under the large irradiances used in our experiment polaritons undergo a significant blueshift, as can be seen in Fig. 2(c)-(g) of the main text. The threshold irradiances for condensation are approximately 1400, 2500 and 3000 kW cm$^{-2}$ for the $P$ flat band, $S$ flat band and AB band respectively. At these thresholds, these three modes remain 0.4, 0.52 and 0.82 meV below the energy of the corresponding cavity modes, which are estimated from the LPB-cavity detuning in the planar region. These values correspond to blueshifts of 37\%, 39\% and 49\% of the LPB-cavity detuning. Beyond these thresholds the condensed modes continue to experience a blueshift with further increase of pumping power, and at the highest irradiance studied (8930 kW cm$^{-2}$) the detunings of the 3 condensates still reside 0.16, 0.33 and 0.39 meV below the cavity. Although the cavity mode may experience a significant redshift under high carrier densities, which could lead to conventional photon lasing being mistaken for polariton lasing \cite{PhysRevB.76.201305}, we believe that in our measurements the emission originates from polaritons, since photon emission is not expected to demonstrate a continuous blueshift with power, responsible for the fragmentation of the flat-band condensates shown in Fig. \ref{frag}. Rather it is expected to show a redshift due to a change in the cavity refractive index.

\end{document}